\documentclass[11pt,letterpaper]{article}
\input setup/preamble
\newtheorem{counter}{Counter}[section]
\newtheorem{theorem}[counter]{Theorem}
\newtheorem{thm}{Theorem}
\newtheorem{lemma}[counter]{Lemma}

\newtheorem{claim}[counter]{Claim}
\newtheorem{fact}[counter]{Fact}

\newtheorem{corollary}[counter]{Corollary}

\newtheorem{definition}[counter]{Definition}

\newcommand{\zo}{\{0,1\}}
\newcommand{\cC}{\mathcal{C}}
\newcommand{\N}{\mathbb{N}}
\renewcommand{\R}{\mathbb{R}}
\renewcommand{\C}{\mathbb{C}}
\newcommand{\cN}{\mathcal{N}}

\newcommand{\cH}{\mathcal{H}}
\newcommand{\cR}{\mathcal{R}}
\newcommand{\sol}{\mathbf{sol}}
\newcommand{\hatp}{\hat{p}}
\newcommand{\hatq}{\hat{q}}
\newcommand{\hatr}{\hat{r}}
\newcommand{\hatA}{\hat{A}}
\newcommand{\hatB}{\hat{B}}
\newcommand{\hatC}{\hat{C}}
\newcommand{\hatu}{\hat{u}}
\newcommand{\hatv}{\hat{v}}
\newcommand{\eps}{\varepsilon}

\newcommand{\nandg}{\mathbf{NAND}}
\newcommand{\matching}{\mathbf{Matching}}
\newcommand{\timp}{\mathbf{2Imp}}

\newcommand{\nand}{\mathbf{Nand}}
\newcommand{\Id}{\mathbf{Id}}

\DeclareMathOperator{\Q}{\{0, 1\}}
\DeclareMathOperator{\aut}{\mathsf{Aut}}
\DeclareMathOperator{\orb}{\mathsf{Orb}}

\newcommand{\abs}[1]{\left\lvert #1 \right\rvert}

\DeclareMathOperator{\size}{\mathsf{Size}}
\crefname{paragraph}{Paragraph}{Paragraphs}
\newcommand{\mohit}[1]{{\color{orange} \footnotesize(Mohit: #1)}}
\newcommand{\navid}[1]{{\color{red} \footnotesize(Navid: #1)}}
\newcommand{\daniel}[1]{{\color{blue} \footnotesize{Daniel: #1}}}
\newcommand{\mohan}[1]{{\color[RGB]{0, 100, 0} \footnotesize(Mohan: #1)}}
\newcommand{\chris}[1]{{\color[RGB]{0, 0, 100} \footnotesize(Chris: #1)}}

\newcommand{\nocomments}{
    \renewcommand{\mohit}[1]{}
    \renewcommand{\navid}[1]{}
    \renewcommand{\daniel}[1]{}
    \renewcommand{\mohan}[1]{}
    \renewcommand{\chris}[1]{}
}

\nocomments % Uncomment to remove all comments from the text

\title{Optimal Depth-Three Circuits for Inner Product} %TODO Please add
\author{Mohit Gurumukhani\thanks{Cornell University, Ithaca, NY, USA. 
Supported by a Sloan Research Fellowship, NSF CAREER Award 2045576, and NSF Award CCF-2514586.
Email: \texttt{mgurumuk@cs.cornell.edu}}
\and
Daniel Kleber\thanks{Department of Computer Science and Engineering, University of California, San Diego. Partially supported by NSF grant 2212136. Email: \texttt{dkleber@ucsd.edu}}
\and
Ramamohan Paturi\thanks{Department of Computer Science and Engineering, University of California, San Diego. Partially supported by NSF grant 2212136. Email: \texttt{rpaturi@ucsd.edu}}
\and
Christopher Rosin \thanks{Constructive Codes, https://constructive.codes.  Email: \texttt{christopher.rosin@gmail.com}}
\and
Navid Talebanfard\thanks{University of Sheffield, Sheffield, UK. Email: \texttt{n.talebanfard@sheffield.ac.uk}}
}
\date{}

\begin{document}
\maketitle

\begin{abstract}

We show that Inner Product in $2n$ variables, $\IP_n(x, y) = x_1y_1 \oplus \ldots \oplus x_ny_n$, can be computed by depth-3 bottom fan-in 2 circuits of size $\poly(n)\cdot (9/5)^n$, matching the lower bound of G\"o\"os, Guan, and Mosnoi (Inform. Comput.'24). Our construction is obtained via the following steps.

\begin{enumerate}
    \item 
    We provide a general template for constructing optimal depth-3 circuits with bottom fan-in $k$ for an arbitrary function $f$. We do this in two steps. First, we partition $f^{-1}(1)$ into \emph{orbits} of its automorphism group. Second, for each orbit, we construct \emph{one} $k$-CNF that (a) accepts the largest number of inputs from that orbit and (b) rejects all inputs rejected by $f$. 

    \item We instantiate the template for $\IP_n$ and $k = 2$. Guided by the intuition (which we call \emph{modularity principle}) that optimal 2-CNFs can be constructed by taking the conjunction of variable-disjoint copies of smaller $2$-CNFs, we use computer search to identify a small set of \emph{building block} 2-CNFs over at most 4 variables.
    
    \item We again use computer search to discover appropriate combinations (disjoint conjunctions) of building blocks to arrive at optimal 2-CNFs and analyze them using techniques from analytic combinatorics.
\end{enumerate}

We believe that the approach outlined in this  paper can be applied to a wide range of functions to determine their depth-3 complexity.

\end{abstract}

\newpage

\section{Introduction}

A $\Sigma_3$ circuit is a depth-3 OR-AND-OR circuit with unbounded fan-in. Despite their simplicity, these circuits are surprisingly powerful. Any $n$-variate Boolean function can be computed by a $\Sigma_3$ circuit of size $O(2^{n/2})$ \cite{Dancik96}, a quadratic speed-up over the trivial construction. More strikingly, Valiant \cite{Valiant77} showed that any linear-size fan-in 2 circuit of logarithmic depth can be computed by a $\Sigma_3$ circuit of size $2^{O(n/\log\log n)}$ size. Therefore, truly exponential lower bounds for depth-3 circuits imply non-linear lower bounds for log-depth circuits. However, the best known depth-3 lower bound for an explicit function is only $2^{\Omega(\sqrt{n})}$ and beating this barrier remains a major open problem in the field \cite{HastadJP95}.

Even a further restriction of depth-3 circuits shows surprising power. Let $\Sigma^k_3$ be the class of $\Sigma_3$ circuits in which every bottom OR gate is connected to at most $k$ input bits. Equivalently, these circuits can be viewed as disjunctions of $k$-CNFs and the number of these formulas serves as a natural measure of size for the circuit. We denote the smallest number of $k$-CNFs needed to express a function $f$ by $\size^k_3(f)$. \footnote{Note that this quantity is off by a factor of $n^k$ from the smallest $\Sigma_3^k$ sized circuit computing $f$. However, since $k$ is small enough in this paper and the circuit size is exponential, this difference is immaterial.} Valiant's depth reduction shows that linear-size log-depth circuits can in fact be computed by $\Sigma^{n^{\epsilon}}_3$ circuits of subexponential size where $\epsilon > 0$ can be chosen arbitrarily. More recently, Golovnev, Kulikov, and Williams \cite{GKW21circuit} showed that \emph{unrestricted} circuits of size $cn$ for every $c < 3.9$ can be computed by $\Sigma^{16}_3$ circuits of size $2^{\epsilon n}$ for some $\epsilon < 1$. Therefore, \emph{near-maximal} lower bounds for $\Sigma^{16}_3$, i.e., lower bounds of the form $2^{n - o(n)}$, imply unrestricted circuit lower bounds beyond the state-of-the-art which is only $3.1n - o(n)$ \cite{FindGHK23,Li022}.

Near-maximal lower bounds are only known for $\Sigma^2_3$ and are due to Paturi, Saks, and Zane \cite{PaturiSZ00}. For $k \ge 3$, the state-of-the-art is due to Paturi, Pudl{\'a}k, Saks, and Zane \cite{PaturiPSZ05} who showed a $\Sigma^k_3$ lower bound of $2^{cn/k}$ where $c > 1$ is a constant, building on a previous work by Paturi, Pudl{\'a}k, and Zane \cite{PaturiPZ99} who proved a tight lower bound of $\Omega(2^{n/k})$ for Parity. 

\paragraph*{Near-maximal lower bound arguments.} Paturi, Saks, and Zane \cite{PaturiSZ00} proved a crucial property of 2-CNFs. They showed that any 2-CNF with $2^{\Omega(n)}$ satisfying assignments, must accept a large \emph{projection}, i.e., an affine space with dimension $\Omega(n)$ defined by equations of the form $x = 0, 1$, $x = y$, and $x = y + 1$. Therefore, any good \emph{affine disperser}, i.e., any function that is not constant under large affine spaces, requires $\Sigma^2_3$ circuits of size $2^{n - o(n)}$. Several explicit constructions of such functions are known including with asymptotically optimal dependence on dimension - $O(\log(n))$ by Li \cite{Li23}. Affine dispersers serve as a natural candidate for near-maximal lower bounds; the known explicit constructions are complicated poly-time algorithms, and the only known circuit construction has super-linear size \cite{HuangIV22}. Furthermore, state-of-the-art unrestricted circuit lower bounds hold for affine dispersers \cite{FindGHK23,Li022}.

In order to prove strong $\Sigma^k_3$ lower bounds for affine dispersers, we cannot hope to apply the projection argument; there are 3-CNF formulas accepting exponentially many assignments which only accept projections of constant dimension \cite{FranklGT22}. However, a more general statement can be true. In particular, we do not know the answer to the following question: \emph{is it true that every $k$-CNF with exponentially many satisfying assignments must accept an affine space of linear dimension?} An affirmative answer yields a near-maximal $\Sigma^k_3$ lower bound for every good affine disperser. However, if the answer is negative, then there may not be a unified proof showing the hardness of \emph{every} affine disperser. Instead, the hardness of each affine disperser may have to be demonstrated in a unique way. This suggests that by determining the complexity of \emph{concrete} affine dispersers, we may be able to develop techniques that will ultimately allow us to prove near-maximal lower bounds.

\paragraph*{Depth-3 complexity of Inner Product.} The Inner Product modulo 2 function $\IP_n$  on $n$ coordinates defined as $ \IP_n(x, y) = x_1y_1 \oplus \cdots \oplus x_ny_n$ serves as an ideal starting point since it has a very simple and concrete definition and it is known to be moderately good affine dispersers (for dimension $n+1$). Frankl, Gryaznov, and Talebanfard \cite{FranklGT22} pointed out that even for $\Sigma^2_3$ circuits, the exact complexity of $\IP_n$ is unclear as the projection technique fails to yield a strong bound. Golovnev, Kulikov, and Williams \cite{GKW21circuit} initiated the study of depth-3 complexity of $\IP_n$ for independent reasons. 

$\IP_n$ can be trivially computed by $\Sigma^2_3$ circuits of size $O(2^n)$: $\IP^1_n = \bigvee_{I \subseteq [n], |I| = \text{odd}} (\wedge_{i \in I}(x_i \wedge y_i) \wedge_{i \not \in I}(\neg x_i \vee \neg y_i))$. A simple lower bound of $\Omega(2^{n/2})$ follows by a reduction from Parity. G\"o\"os, Guan, and Mosnoi \cite{GGM24} showed, somewhat unexpectedly, that both of these bounds can be improved: $(9/5)^n \simeq 2^{0.847n} \le \size^2_3(\IP_n) \le 2^{0.965n}$. They proved the lower bound by identifying a hard probability distribution on inputs to $\IP_n$ and characterizing the formulas that maximize the probability of accepting an input under this restriction. For the upper bound, they used the fomulas in the lower bound to cover solutions with small weight, i.e., the number of coordinates on which both $x$ and $y$ variables are 1 is small. For the remaining solutions, they used a small number of the \emph{matching} formulas. We will discuss these formulas later as we will also use them in our construction. 

Amano \cite{Amano23Majority} improved the upper bound even further to $2^{0.952n}$; 
this was done by observing that a depth-3 circuit for $\IP_n$ can be constructed by partitioning the coordinates into blocks of size $b$ for a small value of $b$, and then combining depth-3 circuits that compute $\IP_b$ and $\neg \IP^b_n$ on these blocks. 
To find such depth-3 circuits for a fixed $b$, the paper used computer search.

\paragraph*{Our results.} In this paper, we determine $\size^2_3(\IP_n)$ up to polynomial factors by showing that the lower bound of \cite{GGM24} is, in fact, tight. 

\begin{thm}[Main result]
\label{thm:main}
$\size^2_3(\IP_n) \le \poly(n)\cdot (9/5)^n$
\end{thm}

The paper makes  two main contributions. Firstly, we give a general framework for the construction of optimal $\Sigma^k_3$ circuits for a large class of 
\emph{highly symmetric} functions.
%, namely highly symmetric functions, i.e., those that remain unchanged under many permutations of their variables. 
We show that it is sufficient to find consistent \emph{extremal} $k$-CNFs to construct $\Sigma^k_3$ optimal circuits for a function $f$. Secondly, our approach to circuit design is somewhat novel: Guided by modularity principle we use computer search  to find the building blocks and  search again for optimal disjoint conjunctions of building blocks to arrive at the desired constructions.
Our methodology is general and flexible and we are optimistic that it can be applied to other functions.

%sales

\section{A recipe for constructing depth-3 circuits}\label{sec:framework}

Let $f : \Q^n \rightarrow \Q$ be an arbitrary Boolean function and consider the task of proving a tight upper bound on $\size_3^k(f)$.
% , the size of the smallest $\Sigma_3^k$ circuit computing $f$. 
In this section, we will provide a general framework for tackling such a task and will instantiate it to construct optimal circuits for $\IP$. 
%We find it valuable to abstract out this framework since it provides a systematic way of constructing near optimal circuits for many classes of Boolean functions and also without this framework, our construction may seem to be constructed out of the blue. 
Our framework reduces the task of constructing $\Sigma_3^k$ circuits for $f$ to the task of constructing many different $k$-CNFs, each \emph{consistent} with $f$, and each one maximizing the number of satisfying assignments with a certain property. We say a $k$-CNF $F$ is consistent with the function $f$ if for all $x\in \zo^n$, it holds that $F(x) = 1 \implies f(x) = 1$, i.e., $\sol(F)\subseteq f^{-1}(1)$, where $\sol(F)$ is the set of satisfying assignments of $F$. We use the term \emph{consistent} $k$-CNF to mean $k$-CNF consistent with a function when the function is clear from the context.
% So, in a $\Sigma_3^k$ circuit for $f$, wire out of the top OR gate goes to a $k$-CNF that must be consistent with $f$.

To describe our framework, we will require some basic concepts from group theory; see 
\cref{subsec:prelimGroupTheory} 
for a quick refresher.
We define the notion of an automorphism group and orbits of a function as follows:

\begin{definition}[Automorphism group and orbits]
For any $f: \zo^n \to \zo$, define the \emph{automorphism group of $f$}, $\aut_f$, to be the group of permutations $\pi$ such that for all $x \in \Q^n$, $f(x_{\pi(1)}, \ldots, x_{\pi(n)}) = f(x_1, \ldots, x_n)$. The action $\aut_f \times f^{-1}(1) \mapsto f^{-1}(1) $ of $\aut_f$ on $f^{-1}(1)$ is defined as follows. $\pi \in \aut_f$ and  $x_1\ldots x_n \in f^{-1}(1)$ maps to  $x_{\pi(1)}\ldots x_{\pi(n)}$. We denote the set of orbits of this action by $\orb_f$. 
\end{definition}
Note that for each $S \in \orb_f$ we have $S \subseteq f^{-1}(1)$. 
We now define the following quantities associated with optimal consistent $k$-CNFs for each orbit. %associated with orbits of $f$ and consistent $k$-CNFs.
\begin{definition}
 For any $f: \zo^n \to \zo$ and $S \in \orb_f$, define 
\begin{align*}
\mu_{f, S, k} := \max_{F: F \text{ is an $n$-variate consistent $k$-CNF}}|\sol(F) \cap S|,\\
\rho_{f, S, k} := |S| / \mu_{f, S, k}, \text{ and } \rho^*_{f, k} = \max_{S \in \orb_f}\rho_{f, S, k}.
\end{align*}
\end{definition}
In words, $\mu_{f, s, k}$ is the largest number of assignments $S$ that are accepted by a consistent $k$-CNF, $\rho_{f, s, k}$ is a lower bound on $\size^k_3(f)$ due to $S$ and $\rho_{f, k}^*$ is the maximum of such lower bounds over all orbits.
We now present the main result of our framework which is proved in \cref{subsec:circuitOrbit}.
%that $\rho^*_{f, k}$ along with the number of orbits, governs the size of optimal $\Sigma_3^k$ circuits for $f$. Formally,
\begin{lemma}
\label{lem:circuitOrbit}
For all $k, n\in \N$ and all functions $f: \zo^n \to \zo$, we have $\rho^*_{f, k} \le \size^k_3(f) \le O\left(n\cdot |\orb_f| \cdot \rho^*_{f,k}\right)$.
\end{lemma}
%We prove this lemma in \cref{subsec:circuitOrbit}.
We note that if  $\abs{\orb_f}$ of a function $f$ is bounded by a polynomial, then $\size^k_3(f)$ and $\rho^*_{f, k}$ are within a polynomial factor of each other. Therefore, to determine the $\Sigma^k_3$ complexity of $f$, it is enough to find a consistent $k$-CNF with the maximum number of solutions in $S$ for every orbit $S$. This is indeed the case for $\IP_n$ and is exactly the strategy we will use to construct optimal $\Sigma_3^2$ circuits for it.

\subsection{Depth-3 upper bound - Proof of \texorpdfstring{\cref{lem:circuitOrbit}}{Lemma 2.3}}
\label{subsec:circuitOrbit}

We will first show that we can cover an entire orbit $S$ by using isomorphic copies of a consistent $k$-CNF $F$ where the number of copies is bounded by $O(n|S|/|\sol(F) \cap S|)$.
%We will need the following claim that shows how we can use a single $k$-CNF consistent with $f$ that covers some fraction of an obit $S$ of $f$ and use it to find not too many $k$-CNFs that completely cover the orbit $S$ of $f$.
\begin{claim}
\label{lem:orbitCover}
Let $f: \zo^n \to \zo$ be an arbitrary function and let $F$ be a $k$-CNF consistent with $f$. Then, for any $S \in \orb_f$, there exist consistent $k$-CNFs $F_1, \ldots, F_t$ such that $S \subseteq \sol(F_1) \cup \ldots \cup \sol(F_t)$ where $t = O(n|S|/|\sol(F) \cap S|)$.
\end{claim}

We first show how the proof of \cref{lem:circuitOrbit} follows from the claim and present the proof of the claim following it. 

\begin{proof}[Proof of \cref{lem:circuitOrbit}]
We prove the lower bound and the upper bound part of the inequality separately.
For the lower bound, fix any $\Sigma_3^k$ circuit $C$ for $f$ .
We express $C = \bigvee_{i = 1}^m F_i$ where $m\in \N$ and each $F_i$ is a $k$-CNF. We will show that for all $S\in \orb_f$, it holds that $m \ge \rho_{f, S, k}$, which implies $m \ge \rho^*_{f, k}$. 
Fix any $S\in \orb_f$.
We observe that each $F_i$ must be consistent with $f$ and so for all $i\in [m]$, we must have that $\abs{\sol(F_i)\cap S} \le \mu_{f, S, k}$. 
Moreover, each assignment in $S$ must be covered by some $F_i$.
Indeed, for each $\alpha\in S$, since $C(\alpha) = f(\alpha) = 1$ and $C = \bigvee_{i=1}^m F_i$, there must exist some $i\in [m]$ such that $F_i(\alpha) = 1$.
Hence, we must have $m \ge |S|/\mu_{f, S, k} = \rho_{f, S, k}$ as desired.

For the upper bound, we construct a circuit $C$ for $f$ as follows.
let $S \in \orb_f$ and let $F_S$ be the $k$-CNF consistent with $f$ that maximizes $\abs{\sol(F_S)\cap S}$.
By definition,  $|\sol(F_S)\cap S| = \mu_{f, S, k}$. 
From \cref{lem:orbitCover} we get consistent $k$-CNFs $F_{S, 1}, \ldots, F_{S, t}$ which cover $S$ where $t = t(S) = O(n|S|/|\sol(F) \cap S|) = O(n \rho_{f, S, k})$.
We define $C = \bigvee_{S\in \orb_f, i\in [t(S)]} F_{S, i}$.
$C$ accepts all assignments from each of the orbits $S$.
Since each $F_{S, i}$ is consistent with $f$, $C$ correctly rejects all assignments from $f^{-1}(0)$ and hence $C$ indeed computes the function $f$.
The total number of $k$-CNFs in $C$  $O(n \sum_{S \in \orb_f} \rho_{f, S, k}) = O(n \cdot |\orb_f| \cdot \rho^*_{f, k})$ as claimed.
\end{proof}

\begin{proof}[Proof of \cref{lem:orbitCover}]
Let $S \in \orb_f$ and $F$ be an $n$-variate $k$-CNF consistent with $f$.
Let $t = 2|S|\ln|S|/|\sol(F) \cap S| = O(n|S|/|\sol(F) \cap S|)$. For each $i \in [t]$, define $F_i = F^{\pi_i}$ to be the permutation of $F$ under a random permutation $\pi_i \in \aut_f$. Fix any $\alpha \in S$. It follows (see \cref{lem:orbitSet} for the general group theoretic claim and its proof) that for a random $\pi \in \aut_f$, $\Pr[\alpha \in \sol(F^{\pi})] = |\sol(F) \cap S|/|S|$, which implies 
\begin{align*}
\Pr[\alpha \not \in \sol(F_1) \cup \ldots \cup \sol(F_t)]
& \le \left(1 - |\sol(F) \cap S|/|S|\right)^t\\
& \le \exp\left(- t\cdot |\sol(F) \cap S|/|S|\right)\\
& < 1 / |S|
\end{align*}
where the last inequality follows by the choice of $t$ and the fact that $\abs{S} \le 2^n$.
By a union bound over all $\alpha\in S$, we conclude that there exists a choice of $F_i$s covering all assignments in $S$.
\end{proof}

\section{Construction for IP}
\label{sec:construction}

In this section, we instantiate our framework from \cref{sec:framework}  to construct  optimal $\Sigma_3^2$ circuits for $\IP$.
We will prove the following main lemma which together with \cref{lem:circuitOrbit} proves \cref{thm:main}. In the following, we also use $\IP^0_n$ for $\IP_n$ and $\IP^1_n = 1-\IP^0_n$ for the complement of $\IP_n^0$.

\begin{lemma}
\label{lem:ip-orbit-cnf} 
There exists a  constant $c$ such that for all integer $n\ge 2$, $\rho^*_{\IP_n, 2} \le n^c \cdot (9/5)^n$.
\end{lemma}
To prove the lemma, we will identify the orbits of $\IP_n$ in \cref{subsec:IPorbits} and the building block $2$-CNFs in \cref{subsec:buildingBlocks}.  In \cref{subsec:infinitelyOften}, we will reduce \cref{lem:ip-orbit-cnf} to a version of itself (\cref{lem:ip-orbit-cnf-inf}) that holds only for infinitely many even $n$ and reduce the constructions for $\IP^1_n$ to those of $\IP^0_{n'}$. Our plan is to construct $2$-CNFs for each of the orbits by taking a \emph{disjoint} conjunction of copies of a small number of the building-block $2$-CNFs. For this purpose, we divide the orbits into six regions in  \cref{subsec:construction-2cnf-orbit}, state \cref{lem:ip-orbit-cnf-inf-region} which is a version of \cref{lem:ip-orbit-cnf-inf} specialized to regions. We prove \cref{lem:ip-orbit-cnf-inf-region} in \cref{sec:construction-2cnf-region}.

% We observe that there is a \emph{systematic} way of finding these $k$-CNFs at least for some functions such as $\IP$.

% \paragraph*{Modularity Principle.}
% We say that a function $f$ satisfies the Modularity Principle if there is a finite set of $k$-CNFs which we call \emph{building blocks} such that for every orbit $S \in \orb_f$, there is a $k$-CNF consistent with $f$ which consists of disjoint copies of building blocks and has maximum number of solutions in $S$.

% Under Modularity Principle, the problem of determining the $\Sigma^k_3$ complexity of a function reduces to finding a finite set of building blocks. Interestingly, for the upper bound problem we do not need to know a prioi whether Modularity Principle holds. As long as we can we construct modular $k$-CNFs consistent with our function, we can try to find basic building blocks to get better and better circuit constructions.

% The rest of the paper follows this plan for $\IP$ to get a $\Sigma^2_3$ circuit upper bound matching the lower bound of \cite{GGM24}.

\subsection{Orbits of IP and spectra of 2-CNFs}\label{subsec:IPorbits} 
We now describe the orbits of $\IP^1_n$ and $\IP^0_n$ and introduce the notion of the \emph{spectrum} of a  $2$-CNF.

We observe that $\aut_{\IP^1_n}= \aut_{\IP^0_n}$ and they contain all permutations that permute the $n$ coordinates, and permute $x_i$ and $y_i$  within each coordinate. 
% We have $|\aut_{\IP^1_n}| = n!2^n$. 
%We now describe the orbits of $\IP_n$.
For input $(x, y)\in (\zo^n)^2$, define the following quantities:
\begin{enumerate}
\item 
$d_2(x, y) = \abs{\{i\in [n]: x_i = y_i = 1\}}$.

\item 
$d_1(x, y) = \abs{\{i\in [n]: x_i\ne y_i\}}$.

\item 
$d_0(x, y) = \abs{\{i\in [n]: x_i = y_i = 0\}}$.
\end{enumerate}
We have $d_2(x, y) + d_1(x, y) + d_0(x, y) = n$.
More importantly, for $j\in \{0,1,2\}$, $d_i(x,y)$ are invariant under a permutation $\pi \in \aut_{\IP^1_n}$, that is, $d_j(x', y') = d_j(x, y)$ where $(x',y')=\pi(x,y)$.
We also observe that if $d_j(x, y) = d_j(x', y')$ for all $j\in \{0, 1, 2\}$,  then there exists a permutation $\pi\in \aut_{\IP^1_n}$ such that $\pi(x, y) = (x', y')$.
Hence, $d_j$ precisely determine the orbits of $\IP^1_n$ and $\IP^0_n$
In particular we have $|\orb_{\IP^1_n}| = |\orb_{\IP^0_n}|= O(n^2)$.
We parameterize the orbits of $\IP_n$ as follows:
\begin{definition}[Orbits of $\IP^1_n$ and $\IP^0_n$]
For $d_2, d_1, d_0\in [n]$ such that $d_2 + d_1 + d_0 = n$, let $S(d_2,d_1,d_0)\subseteq (\zo^n)^2$ denote the set of assignments $(x,y) \in (\zo^n)^2$ such that $d_j(x,y) = d_j$ for all $j\in \{0, 1, 2\}$. Note that $S(d_2,d_1,d_0)$ is an orbit of the function $\IP^{d_2 \mod 2}_n$ and for any $(x,y)\in S(d_2,d_1,d_0)$,
$\IP^0_n(x,y)=  d_2 \mod 2$.
\end{definition}
We record the following fact regarding size of each of the orbits:

\begin{fact}
\label{fact:IPOrbitSize}
$|S(d_2,d_1, d_0)| = \binom{n}{d_2} \binom{n-d_2}{d_0}2^{n-d_2-d_0} = \binom{n}{d_2} \binom{d_1+d_0}{d_1}2^{d_1}$.
\end{fact}

We now define the spectrum of a $2$-CNF $F$ as the generating function that encodes the number of its satisfying assignments from each of the orbits. 
%one more quantity associated with a $2$-CNF $F$ that will capture how its satisfying assignments intersect various orbits of $\IP_n$. We will do this by associating a generating function to each such $F$ where the  monomials represent an orbit of $\IP_n$ and the coefficient for the monomials will represent the number of satisfying assignments of $F$ from that orbit. Formally,
\begin{definition}[Spectrum of 2-CNF]
For a $2$-CNF $F$ over $2n$ variables, define the spectrum of $F$ to be the generating function $G(x, y, z) = \sum_{p, q, r\in \N} C_{p, q, r} x^p y^q z^r$ where $C_{p, q, r}$ is the number of assignments $\alpha\in \zo^{2n}$ such that $F(\alpha) = 1$, $d_2(\alpha) = p, d_1(\alpha) = q$, and $d_0(\alpha) = r$.
\end{definition}
The spectrum of $2$-CNF of $F$ is always a finite homogeneous polynomial. If $F$ is consistent with  $\IP^0_n$, then $C_{p, q, r}=0$ if $(p,q,r)$ is not an orbit of $\IP^0_n$.

\subsection{Building blocks for our constructions}\label{subsec:buildingBlocks}

% To prove \cref{lem:ip-orbit-cnf}
%To construct consistent $2$-CNFs maximizing the number of solutions from a fixed orbit, 
In this section, we identify a small number of building block 2-CNFs which will be used to construct $2$-CNFs; we will also define the notion of a disjoint conjunction of the building blocks. Since any function over $2$ variables can expressed as a $2$-CNF, our building block descriptions will be expressed as $\mathrm{AND}$ of functions over $2$ variables. 

\begin{definition}[Building block $2$-CNFs and their spectra]\label{def:buildingBlocks}
%We define the following $2$-CNFs that we will refer to as building block $2$-CNFs:
We define the following building blocks:
\begin{itemize}
\item 
$\Id_2$
denotes the $2$-CNF over $1$ coordinate ($2$ variables) that accepts iff both variables equal $1$; its spectrum is $x$.

\item 
$\Id_1$ denotes the $2$-CNF over $1$ coordinate that accepts iff both variables do not equal each other; its spectrum is $2y$.

\item 
$\Id_0$ denotes the $2$-CNF over $1$ coordinate that accepts iff both variables equal $0$; its spectrum is $z$.

\item
$\nandg$ denotes the $2$-CNF over $1$ coordinate that accepts iff both variables do not equal $1$ ; its spectrum is $2y+z$.         

\item
$\matching$ denotes the $2$-CNF over over $2$ coordinates defined as follows.
On input $(x, y)\in (\zo^2)^2$, the $2$-CNF accepts iff $x_1 = x_2$, and $y_1 = y_2$ ; its spectrum is $x^2 + 2y^2 + z^2$.

\item
$\timp$ denotes the $2$-CNF over $2$ coordinates defined as follows.
On input $(x, y)\in (\zo^2)^2$, the $2$-CNF accepts iff the following holds: $x_1 = x_2$, $x_1 \implies y_1$, and $x_2 \implies y_2$ ; its spectrum is $x^2 + y^2 + 2yz + z^2 =  x^2 + (y + z)^2$.
\end{itemize}
\end{definition}

\begin{definition}[Disjoint conjunction]
\label{def:disjointConjunction}
Let $F_1$ and $F_2$ be arbitrary $2$-CNFs over $n_1$ and $n_2$ variables respectively.
Then, $F_1\land F_2$ is the $2$-CNF over $n_1 + n_2$ variables obtained by adding all clauses from both $F_1$ and $F_2$ where we associate the first $n_1$ variables with variables of $F_1$ and the last $n_2$ variables with variables of $F_2$.
\end{definition}
%When we apply disjoint conjunction over more than $2$ $2$-CNFs, since the order of the conjunction does not matter, we apply it in some canonical order. 

We next record a couple of crucial observations regarding the spectrum of the disjoint conjunction of  $2$-CNFs and $2$-CNFs consistent with either $\IP^0_n$ or $\IP^1_n$.
\begin{fact}[Spectra of disjoint conjunctions $2$-CNFs]\label{fact:spectraOfDisjointCNFs}
Let $F_1$ and $F_2$ be arbitrary $2$-CNFs over $n_1$ and $n_2$ variables with spectra $G_1$ and $G_2$ respectively.
Then, the spectrum of $F_1\land F_2$ equals $G_1\cdot G_2$.
Furthermore, if $F_1$ is consistent with $\IP_{n_1}^{b_1}$ and that $F_2$ is consistent with $\IP_{n_2}^{b_2}$ for some $b_1, b_2\in \zo$, then
$F_1\land F_2$ is consistent with $\IP_{n_1 + n_2}^{b_1 \oplus b_2}$.
\end{fact}

%For $b\in \zo$, we define the function $\IP_n^b$ to equal $\IP_n$ if $b = 1$ and to equal $\neg \IP_n$ if $b = 0$. 
%For $b\in \zo$, a $2$-CNF $F$ is consistent with $\IP_n^b$  if and only if the following holds: for each monomial $x^py^qz^r$ in the spectrum of $F$,   if its coefficient is non-zero, then $p\pmod 2 = b$.
%We now have the following fact regarding consistency of disjoint conjunctions of consistent $2$-CNFs:

%\begin{fact}
%\label{fact:spectraOfDisjointCNFs}
%Let $F_1$ and $F_2$ be $2$-CNFs on $n_1$ and $n_2$ variables respectively.
%Furthermore, assume that there exist $b_1, b_2\in \zo$ such that $F_1$ is consistent %with $\IP_{n_1}^{b_1}$ and that $F_2$ is consistent with $\IP_{n_2}^{b_2}$. Then,
%$F_1\land F_2$ is consistent with $\IP_{n_1 + n_2}^{b_1 \oplus b_2}$.
%\end{fact}

We note that each building block from \cref{subsec:buildingBlocks} is consistent with $\IP_n^b$ for some $b\in \zo$ for an appropriate $n$. This implies:

\begin{corollary}\label{cor:consistencyOfBuildingBlocks}
Let $F$ be a $2$-CNF over $n$ variables obtained by (repeatedly) applying disjoint conjunction to building blocks from \cref{def:buildingBlocks}. Then, $F$ is consistent with $\IP_n^b$ for some $b\in \zo$.
\end{corollary}

\subsection{Reducing to infinitely often case}\label{subsec:infinitelyOften}

% We now show how to appropriately combine $2$-CNFs to obtain consistent $2$-CNFs that accept many assignments from an orbit of $\IP_n$ and prove \cref{lem:ip-orbit-cnf}. 
We now state a version of  \cref{lem:ip-orbit-cnf} that only holds for infinitely many $n$, each a multiple of some even  constant.  
\begin{lemma}
\label{lem:ip-orbit-cnf-inf}
There exist  constants $c$, even $m\in \N$ such that the following holds:
For all $p, q, r, n\in \N$ such that $n = p + q + r$ and $m$ divides each of $p, q, r$ and $n$, there exists a $2$-CNF $F$ over $n$ variables that is consistent with $\IP_n^0$ and satisfies:
\[
\frac{\abs{S(p, q, r)}}{\abs{\sol(F)\cap S(p, q, r)}} \le n^c\cdot (9 / 5)^n.
\]
\end{lemma}

We will now show that \cref{lem:ip-orbit-cnf} follows from \cref{lem:ip-orbit-cnf-inf}.
\cref{lem:ip-orbit-cnf-inf} will be proven by reducing it to \cref{lem:ip-orbit-cnf-inf-region} in \cref{subsec:construction-2cnf-orbit}.

\begin{proof}[Proof of \cref{lem:ip-orbit-cnf}]

Fix any integer $n\ge 2$ and let $c_0$ and $m$ be as given by \cref{lem:ip-orbit-cnf-inf}.
Now consider any orbit $S$ of $\IP_n$.
It suffices to show that there exists a $2$-CNF $F$ consistent with $\IP_n$ such that $\frac{\abs{S}}{\abs{\sol(F)\cap S}} \le n^{c}\cdot (9 / 5)^n$ where $c$ is a  constant.
As discussed in \cref{subsec:IPorbits}, $S$ is parameterized by integers $p, q, r\in \N$ such that $p + q + r = n$ where $p$ is odd and $S = S(p, q, r)$.

Let $p'\le p, q'\le q, r'\le r$ be the largest multiples of $m$ less than or equal to $p, q, r$ respectively. 
Let $n' = p' + q' + r'$.
$n'$ must also be a multiple of $m$ and $n - n' \le 3m$.
We apply \cref{lem:ip-orbit-cnf-inf} for  $p', q', r', n'$ to obtain $2$-CNF $F'$ over $n'$ variables such that $\frac{\abs{S(p', q', r')}}{\abs{\sol(F')\cap S(p', q', r')}} \le (n')^{c_0}\cdot (9 / 5)^{n'}$.
Let $F$ be the $2$-CNF over $n$ variables obtained by the disjoint conjunction of $p - p'$ copies of $\Id_2$, $(q - q')$ copies of $\Id_1$, $(r - r')$ copies of $\Id_0$, and one copy of $F'$.

We know that $F'$ is consistent with $\IP_{n'}^{p'\pmod 2}$, $\Id_1$ and $\Id_0$ are consistent with $\IP_1^0$ and $\Id_2$ is consistent with $\IP_1^1$. Applying \cref{fact:spectraOfDisjointCNFs}, we get that $F$ is consistent with $\IP_n^{p\pmod 2}$ as desired.
Next, using \cref{fact:IPOrbitSize} and the fact that $n - n' \le 3m, p-p'\le m, q-q'\le m, r-r'\le m$, we infer that there exists a  constant $c_1$ such that $\frac{\abs{S(p, q, r)}}{\abs{S(p', q', r')}} \le n^{c_1}$.
We also have $\abs{\sol(F)\cap S(p, q, r)} = 2^{q - q'}\abs{\sol(F')\cap S(p', q', r')}$.
We thus have 
\begin{align*}
\frac{\abs{S(p, q, r)}}{\abs{\sol(F)\cap S(p, q, r)}} 
& = \frac{\abs{S(p, q, r)}}{\abs{S(p', q', r')}}\cdot \frac{\abs{S(p', q', r')}}{\abs{\sol(F)\cap S(p, q, r)}}\\
& = 2^{q' - q}\cdot \frac{\abs{S(p, q, r)}}{\abs{S(p', q', r')}}\cdot \frac{\abs{S(p', q', r')}}{\abs{\sol(F')\cap S(p', q', r')}}\\
& \le 2^m\cdot n^{c_1}\cdot n^{c_0}\cdot (9 / 5)^n\\
& \le n^c\cdot (9 / 5)^n
\end{align*}
for a suitable constant $c$.
\end{proof}

\subsection{Splitting orbits into regions}\label{subsec:construction-2cnf-orbit}

In this subsection, we divide the orbits of $\IP^1_n$ and $\IP^0_n$  into six regions and  prove \cref{lem:ip-orbit-cnf-inf} for each of the six regions.

\begin{definition}\label{def:regionOrbitals}
For $i\in [6]$ and $n\in \N$, let $\cR_i(n)\subseteq \N\times \N\times \N$ is a set of triples $(p, q, r)\in \N$ such that $p + q + r = n$ and 
%the tripe $(p,q,r)$ satisfies the following constraints (which are presented for each region).
\begin{enumerate}
\item $\cR_1(n)$: $\frac{1}{2}n - \frac{25}{32}p \ge 0$, $\frac{5}{4}n - \frac{25}{32}p - \frac{25}{4}r \ge 0$, and $-\frac{5}{4}n + \frac{25}{16}p + \frac{25}{4}r \ge 0$.
%The triples in $\cR_1(n)$ satisfy the constraints:
%\begin{itemize}
%    \item 
%    $\frac{1}{2}n - \frac{25}{32}p \ge 0$.
%    
%    \item 
%    $\frac{5}{4}n - \frac{25}{32}p - \frac{25}{4}r \ge 0$.
%    
%    \item 
%    $-\frac{5}{4}n + \frac{25}{16}p + \frac{25}{4}r \ge 0$.
%\end{itemize}

\item $\cR_2(n):$ $\frac{1}{2}n - \frac{25}{32}p \le 0$.
%The triples in $\cR_2(n)$ satisfy the constraints:
%\begin{itemize}
%    \item 
%    $\frac{1}{2}n - \frac{25}{32}p \le 0$.
%\end{itemize}

\item $\cR_3(n):$ $\frac{1}{2}n - \frac{25}{32}p \ge 0$, $\frac{5}{4}n - \frac{25}{32}p - \frac{25}{4}r \le 0$, and $\frac{1}{4}n - p \le 0$.

\begin{comment}  

The triples in $\cR_3(n)$ satisfy the constraints:
\begin{itemize}
    \item 
    $\frac{1}{2}n - \frac{25}{32}p \ge 0$.

    \item 
    $\frac{5}{4}n - \frac{25}{32}p - \frac{25}{4}r \le 0$.
    
    \item 
    $\frac{1}{4}n - p \le 0$.
\end{itemize}
\end{comment}

\item $\cR_4(n):$ $\frac{1}{2}n - \frac{25}{32}p \ge 0$, $-\frac{5}{4}n + \frac{25}{16}p + \frac{25}{4}r \le 0$, $-\frac{1}{20}n + r \ge 0$, and $\frac{1}{4}n - p \le 0$.

\begin{comment}
The triples in $\cR_4(n)$ satisfy the constraints:
\begin{itemize}
    \item 
    $\frac{1}{2}n - \frac{25}{32}p \ge 0$.

    \item 
    $-\frac{5}{4}n + \frac{25}{16}p + \frac{25}{4}r \le 0$.

    \item 
    $-\frac{1}{20}n + r \ge 0$.
    
    \item 
    $\frac{1}{4}n - p \le 0$.
\end{itemize}
\end{comment}

\item $\cR_5(n):$ $\frac{1}{2}n - \frac{25}{32}p \ge 0$, and $-\frac{1}{20}n + r\le 0$.
\begin{comment}
The triples in $\cR_5(n)$ satisfy the constraints:
\begin{itemize}
    \item 
    $\frac{1}{2}n - \frac{25}{32}p \ge 0$.

    \item 
    $-\frac{1}{20}n + r\le 0$.
\end{itemize}
\end{comment}
\item $\cR_6(n):$ $\frac{1}{4}n - p \ge 0$.
%The triples in $\cR_6(n)$ satisfy the constraints:
%\begin{itemize}
%   \item 
%    $\frac{1}{4}n - p \ge 0$.
%\end{itemize}
\end{enumerate}
\end{definition}

From here on, we refer to  $(p,q,r)$ as the orbit corresponding to it.

For each region $\cR_i$ and each orbit in the region, we will construct a $2$-CNF such that its spectrum satisfies the following property. 

\begin{lemma}
\label{lem:ip-orbit-cnf-inf-region}
For all $i\in [6]$, there exist  constants $K_i$ and even $m_i\in \N$ such that the following holds:
For all $p, q, r, n\in \N$ such that $(p, q, r)\in \cR_i$, $n = p + q + r$, and $m_i$ divides each of $p, q, r$,    there exists a $2$-CNF $F$ over $2n$ variables that is consistent with $\IP_n^{0}$ and satisfies:
\[
\frac{\abs{S(p, q, r)}}{\abs{\sol(F)\cap S(p, q, r)}} \le n^{K_i}\cdot (9 / 5)^n.
\]
\end{lemma}

We first show that \cref{lem:ip-orbit-cnf-inf} follows from  \cref{lem:ip-orbit-cnf-inf-region}.
We will prove \cref{lem:ip-orbit-cnf-inf-region}  in \cref{sec:construction-2cnf-region}.

\begin{proof}[Proof of \cref{lem:ip-orbit-cnf-inf}]
We apply \cref{lem:ip-orbit-cnf-inf-region} and obtain even $m_i$ and $K_i$. Let $m$ be the least common multiple of $m_1, m_2, m_3, m_4, m_5,$ and $m_6$ and let $c$ to be the maximum of $K_1, K_2, K_3, K_4, K_5, K_6$. We  observe that the six regions cover the set of all triples such that $p + q + r = n$, completing our proof.
\end{proof}

\section{An overview of constructions}

In this section we provide an overview of our constructions that will lead to the proof of
\cref{lem:ip-orbit-cnf-inf-region}.

%\paragraph{Setup and Common analysis}
%The general setup is that for given $p, q, r$ from some region $\cR_i$, we want to construct a $2$-CNF $F$ such that $\frac{\abs{\sol(F)\cap S(p, q, r)}}{\abs{S(p, q, r}} \le \poly(n)\cdot (9 / 5)^n$ where $n = p + q + r$ and where $F$ is consistent with $\IP_n^{p\pmod 2}$. We will ensure that $p$ is even so that this reduces to showing that $F$ is consistent with $\IP_n^0$.

 For each $i\in [6]$ and $(p, q, r) \in \cR_i$, we construct a $2$-CNF by taking a disjoint conjunction of an appropriate number of building block $2$-CNFs while ensuring that $F$ is consistent with $\IP_n^0$. We then argue  that $F$ captures many assignments from $S(p, q, r)$ by lower bounding  the coefficient of the monomial $x^py^qz^r$ in its spectrum. This is the common framework of construction and analysis. To illustrate  how the framework works, we  provide a sketch of the  construction and its analysis for region 1 in the following. Our analysis of the constructions will vary in their techniques across regions, ranging from direct calculations to reducing to an optimization problem that we solve using a numerical optimization software (see \cref{subsubsec:2CNFRegion3} for an example). For a lot of regions, we will use tools from multivariate analytic combinatorics to lower bound the coefficient of the desired monomial.
 
 A more complete description of the constructions for each region and their analysis is presented in \cref{sec:construction-2cnf-region} and \cref{sec:coeffExtract}.

\paragraph{Sketch of the construction and analysis for $\cR_1$.}
%We here sketch the proof of \cref{lem:ip-orbit-cnf-inf} for $i=1$.

Let $A = \frac{5}{4}n - \frac{25}{32}p - \frac{25}{4}r, B = \frac{-5}{4}n + \frac{25}{16}p + \frac{25}{4}r, C = \frac{1}{2}n - \frac{25}{32}p$. We select $m_1 = 8000$ (which is not necessarily optimized) and require that $m_1$ divides $p$, $q$, and $r$. 
$A, B, C$ are non-negative integers since  $(p, q, r) \in \cR_1$.
We take a disjoint conjunction of $A$ copies of $\matching$, $B$ copies of $\timp$ and $2C$ copies of $\nand$ building blocks to construct our $2$-CNF $F$. Since $m_1$ is even, we have that $F$ is consistent with $\IP_n^0$.

%We select $m_1 = 8000$ (for example) and require that $m_1$ divides $p$, $q$, and $r$. 
 
%by requiring that each of $A, B, C$ is even by imposing divisibility requirements on $p, q, r$ (by setting $m = 8000$ say).
The spectrum of $F$ equals $P(x, y, z) = (x^2 + 2y^2 + z^2)^A (x^2 + (y + z)^2)^B ((2y+z)^2)^C$ and the coefficient of $x^py^qz^r$ in $P(x,y,z)$ equals $\abs{\sol(F)\cap S(p, q, r)}$.  
To bound the coefficient, we define $f(u, v) = P(u^{1/2}, v, 1)$.
Since $p + q + r = n$ and $P$ is a homogeneous polynomial, the coefficient of $u^{p/2}v^q$ in $f(u, v)$ equals the coefficient of $x^py^qz^r$ in $P(x, y, z)$.
We set the gradient of $f(u, v)$ to $0$ and compute the non-negative real solutions of this equation to get the unique \emph{critical point} at $(16, 2)$.
We then apply a standard result in analytic combinatorics - \cref{lem:bivariateOptApproxCauchy} - that states that  for \emph{well-behaved} $f$, the value of the coefficient of $u^{p/2}v^q$ (up to a polynomial factor) equals  $\frac{f(16, 2)}{16^{p/2}2^{q}} = \frac{25^{A+B+C}}{16^{p/2}2^{q}} = \frac{5^n}{2^{2p+q}}$ where we used the fact that $A + B + C = n/2$. 
%See \cref{lem:coeffBoundIntegerCase} for details.
We now have $\abs{\sol(F) \cap S(p, q, r)} = \frac{5^n}{2^{2p+q}}$ up to a polynomial factor. 
We upper bound the number of 2-CNFs required to cover the orbit $S(p,q,r)$.  
\begin{align*}
\frac{\abs{S(p, q, r)}}{\abs{\sol(F)\cap S(p, q, r)}} 
& \le \binom{n}{p}\binom{n-p}{r}\cdot 2^{n - p - r}\cdot \frac{2^{2p+q}}{5^n}\\
& = \binom{n}{p, q, r}\cdot 4^{-r}\cdot(4/5)^n\\
& \le  \sum_{i, j, k\in \N: i + j + k = n} \binom{n}{i, j, k}\cdot 1^i \cdot 1^j\cdot 4^{-k}\cdot (4 / 5)^n\\
& = (9 / 4)^n \cdot (4 / 5)^n  = (9 / 5)^n
\end{align*}   
where we used \cref{fact:IPOrbitSize} to establish the first inequality.
%binomial theorem in the second last line. 
%Hence, our $F$ has the desired properties.

\begin{comment}

\paragraph{Organization} The rest of the paper is organized as follows: In \cref{sec: conclusion}, we provide conclusion and discussion; in \cref{sec: preliminaries}, we setup preliminaries and notation; in \cref{sec:construction-2cnf-region}, we provide constructions for each region, relying on bounds on coefficients of appropriate monomials of generating functions; in \cref{sec: coeff extract} we prove these required bounds on coefficients using  a result in analytic combinatorics; in \cref{subsec:bivariateOptApproxCauchy} we prove the exact result from that we need using basic results in analytic combinatorics (even though the result we require seems standard in the area, we could not find it in the literature in the precise form that we needed).
\end{comment}

\section{Conclusions}
\label{sec: conclusion}

%\subsection{Open Questions}
The immediate question left open is to determine $\size^k_3(\IP_n)$ for every $k \ge 3$. We believe that a similar methodology should work for upper bounds. However, the challenge is to find an argument that scales with $k$. It also remains open to find a non-trivial lower bound argument for $k\ge 3$. 

Another related open question is to construct an explicit degree $2$ polynomial that requires $\Sigma_3^k$ circuits of size $2^{n - o(n)}$ for all constant $k$; this is open even for $k = 2$. Existentially, Impagliazzo, Paturi, and Zane showed that a random degree $2$ polynomial requires maximal $\Sigma_3^k$ circuits \cite{IPZ01exponential}.

\subsection{How did we discover these circuits?}

We here document our journey for constructing optimal circuits for $\IP_n$. We hope this inspires future work to find optimal or close to optimal $\Sigma_3^k$ circuits for various explicit functions.

\paragraph{Modular constructions}
As a starting point, we believed that optimal circuits for $\IP_n$ must be modular, i.e., obtained by combining disjoint copies of smaller $2$-CNFs. Our reasons for such a belief are twofold: All previous best known constructions of $\IP_n$ had this property and  optimal (and conjectured to be optimal) $\Sigma_3^k$ circuits for other simple functions such as Parity and Majority have this property.

\paragraph{Searching for better circuits for small $n$.} Our first improvement came from examining the construction of $\Sigma_3^2$ circuits for $\IP_n$ by Amano \cite{Amano23Majority}. 
% who showed that $\IP_n$ can be computed by a disjunction of at most $2^{cn}$ 2-CNF formulas, where $c \approx 0.9518$.
To construct such circuits, the paper used an integer programming solver on a set of constraints to find a disjunction of 7 $2$-CNFs that compute $\IP^1_4$ and another disjunction of 7 $2$-CNFs that compute $\IP^0_4$. For the general construction, a combination of these formulas is used. 

We searched for better circuits using Kissat SAT solver under a similar set of constraints and identified a disjunction of 14 $2$-CNF formulas that computes $\IP^1_5$ and another disjunction of 13 $2$-CNF formulas that computes $\IP^0_5$. Repeated use of these formulas as in Amano's construction gives a slightly improved construction for $\IP_n$ with size $2^{cn}$ where $c \approx 0.9509$. This inspired us to seek further improvements.

\paragraph{Search for small Pareto-optimal building blocks}

Suppose $A$ and $B$ are $2$-CNFs consistent with $\IP^b_n$ for some $b\in \zo$ and $n\in \N$.
Then, we say $A$ dominates $B$ if each coefficient in $A$'s spectrum is at least as large as the corresponding coefficient in $B$'s spectrum. 
We performed an exhaustive computer search for Pareto-optimal (non-dominated) 2-CNF building blocks for $\IP^0_2$ that identified three distinct Pareto-optimal building blocks: $\matching, \timp$ and disjoint conjunction of 2 copies of $\nand$. In particular, $\timp$ was a new building block first identified by this exhaustive search and is \emph{crucial} in our optimal constructions. We also identified other Pareto-optimal building blocks for small values of $n$ that we did not end up using.

\paragraph{Search for compositions of building blocks using disjoint conjunctions}
After we identified a suitable set of building blocks, we are left with the task of finding disjoint conjunctions of the building blocks to cover the assignments in each orbit. This task turned out to be highly nontrivial and we resorted to computer search again to gain intuition to guide us towards optimal general constructions.  We searched for efficient disjoint conjunctions of building blocks for each orbit of $\IP_n$ for some large fixed values of $n$.
For this, we enumerated all distinct compositions of the building blocks. This exercise pointed to the best composition for each orbit and helped identify the hardest orbits. 

For the search, we used  $n=50$, $n=100$, and $n=200$. We have also enumerated restricted compositions for  $n=400$ and $n=800$ near the conjectured hardest orbits.
The search identified best compositions involving the building blocks $\matching, \timp$, and $\nand$ which led to the general constructions. From our best compositions for each value of $n$, we obtained depth-3 circuits of size $\poly(n) \cdot 2^{cn}$ for $\IP_n$ where the dependence of $c$ on $n$ is presented below.
\begin{table}[h!]
    \centering
    \begin{tabular}{cc}
        $n=50$ & $c=0.8344320$\\
        $n=100$ & $c=0.8414042$\\
        $n=200$ & $c=0.8447647$\\
        $n=400$ & $c=0.8463819$\\
        $n=800$ & $c=0.8471913$ \\
    \end{tabular}
\end{table}

A least-squares fit to the data yields $0.8479999-0.6470509/n$, suggesting an eventual convergence to $0.8479999$ for large $n$.  This is extremely close to $\log(9/5)=0.8479969$, which would match the lower bound from \cite{GGM24}, motivating us to pursue an upper bound with $c=\log(9/5)$.

\paragraph{Restricted compositions to cover regions}
Finally, we analytically solved for optimal constructions using saddle point methods with  $\matching, \timp$, and $\nand$  as building blocks. Our analysis identified regions 2, 3, and 4 and that optimal constructions for these regions use at most $1$ or $2$ building blocks.
We then performed a search for optimal compositions (using at most 2 types of building blocks) for each of the regions for $n=200$. 
The goal of the search was to find a small number of distinct compositions that would cover a region so that the ratio of the size of an orbit to number of solutions in the orbit is at most $(9 / 5)^n$.
This search successfully identified  the desired compositions that are sufficient for our purpose.

% \section{Notation and facts from group theory}
\section{Preliminaries}
\label{sec:preliminaries}

We use $\log(x)$ to denote logarithm of $x$ with base $2$ and $\ln(x)$ to denote the natural logarithm. The  binary entropy function $H: [0, 1] \to [0, 1]$ is defined as $H(x) = x\log(1/x) + (1-x)\log(1 / (1-x))$.

We will utilize the following well known approximation for binomial coefficients (see, for instance, equation 7.14 from \cite{tc1999infotheory}):
\begin{theorem}\label{thm:binomialApprox}
For all $m, k\in \N$ such that $m\ge k$, we have that
\[
    \frac{1}{m+1}2^{m H(k / m)} \le \binom{m}{k}\le 2^{m H(k / m)}.
\]
%where $H$ is the binary entropy function.
\end{theorem}

\subsection{Group theory}
\label{subsec:prelimGroupTheory}

%We here define and prove some basic concepts from group theory.

\begin{definition}[Group action and orbits]
An action of a finite group $G$ on a finite set $S$ is a mapping $(\cdot)$ from $G \times S$ to $S$ such that
\begin{itemize}
    \item $e \cdot x = x$ for all $x \in S$ where $e$ is the identity element of $G$,
    \item $g \cdot (h \cdot x) = gh \cdot x$ for all $g,h \in G$ and $x \in S$.
\end{itemize}
The orbit of an element $x \in S$ is $G \cdot x = \{g \cdot x : g \in G\}$. The properties of the group show that orbits partition $S$ into natural equivalence classes.
\end{definition}

For $g \in G$ and $T \subseteq S$ define $g \cdot T = \{g \cdot y : y \in T\}$. For $x \in S$ let the stabilizer subgroup of $x$ be defined as $G_x = \{g \in G : g\cdot x = x\}$.

\begin{theorem}[Orbit-Stabilizer Theorem]
For every $x \in S$, $|G\cdot x| = |G|/|G_x|$.
\end{theorem}

We need the following lemma which is an easily corollary of Orbit-Stabilizer Theorem.

\begin{lemma}
\label{lem:orbitSet}
Let $x \in S$ and let $T \subseteq S$. Then $\Pr_g[x \in g \cdot T] = \frac{|G \cdot x \cap T|}{|G \cdot x|}$, where $g$ is sampled uniformly from $G$.
\end{lemma}

\begin{proof}

we have $\Pr_g[x \in g \cdot T] = \Pr_g[\exists y \in T: x = g \cdot y]$. The events $x = g \cdot y$ are disjoint and have non-zero probability only when $y$ is in the same orbit as $x$. Therefore we can write the probability as $\sum_{y \in G \cdot x \cap T}\Pr_g[x = g \cdot x]$. It remains to show that $\Pr_g[x = g \cdot y] = 1/|G \cdot x|$ for every $y \in G \cdot x \cap T$. 

Observe that there is a bijection between $A := \{g : g \cdot y = x\}$ and $G_y$. Pick any $\sigma \in A$. Then the mapping $\pi$ to $\sigma^{-1}\pi$ is a bijection. 
Therefore $\Pr_g[x = g \cdot y] = |A|/|G| = |G_y|/|G| = 1/|G \cdot x|$, where the last equality follows from Orbit-Stabilizer Theorem and the fact that $y$ and $x$ are in the same orbit. 
\end{proof}

\section{Constructing 2-CNFs for each region}\label{sec:construction-2cnf-region}

In this section, we will prove \cref{lem:ip-orbit-cnf-inf-region} for each of the regions by constructing $2$-CNFs for orbits in that region.
%go over each of the $6$ regions (in each subsequent subsection), as described in \cref{def:regionOrbitals} and prove \cref{lem:ip-orbit-cnf-inf-region} for each of these regions
Proofs provided in this section depend on results regarding asymptotes of coefficients of multivariate generating functions which are proved in  \cref{sec:coeffExtract}.

\subsection{2-CNFs for orbits in region 1}\label{subsubsec:construction-2cnf-orbit-region-1}
In this subsection, we will prove \cref{lem:ip-orbit-cnf-inf-region} for orbits in $\cR_1$.
We will rely on the following result regarding asymptotes of a coefficient of a monomial from a generating function, which we will prove in \cref{subsec:R1CoeffExtract}.
\begin{lemma}
\label{lem:coeffBoundIntegerCase}
There exists a  constant $C_0$ such that the following holds.
Let $A, B, C, p, q, r, n$ be nonnegative integers such that
$n\ge C_0$, $n = p + q + r, A = \frac{5}{4}n - \frac{25}{32}p - \frac{25}{4}r, B = \frac{-5}{4}n + \frac{25}{16}p + \frac{25}{4}r, C = \frac{1}{2}n - \frac{25}{32}p$.
Then, the coefficient of the monomial $x^p y^q z^r$ in $P(x, y, z) = (x^2 + 2y^2 + z^2)^A (x^2 + (y + z)^2)^B ((2y+z)^2)^C$ is at least $\frac{1}{n^{C_0}} \frac{5^n}{2^{2p + q}}$.
\end{lemma}

%Let's see how using this, our desired claim easily follows:
\begin{proof}[Proof of \cref{lem:ip-orbit-cnf-inf-region} for region 1]
Let $m_1 = 32$.
Now consider any $(p, q, r)\in \cR_1$ such that $m_1$ divides each of $p$, $q$, and $r$ and hence $n$.
%We will construct $2$-CNF $F$ that is consistent with $\IP_n^{p\pmod 2}$ and such that $\frac{\abs{S(p, q, r)}}{\abs{\sol(F)\cap S(p, q, r)}} \le n^{C_0}\cdot(9 / 5)^n$ for some  constant $K_0$ (that is also independent of the choice of $p, q, r$).

Let $A = \frac{5}{4}n - \frac{25}{32}p - \frac{25}{4}r, B = \frac{-5}{4}n + \frac{25}{16}p + \frac{25}{4}r, C = \frac{1}{2}n - \frac{25}{32}p$.
$A, B, C$ are non-negative integers since
$m_1$ divides each of $n, p, q, r$ and by the definition of $\cR_1$ we have that $A, B, C\ge 0$.

Our $2$-CNF $F$ is  obtained by a disjoint conjunction of $A$ copies of $\matching$, $B$ copies of $\timp$ and $2C$ copies of $\nand$. It is easy to see that  $F$ is consistent with $\IP^{p\mod 2}_n = \IP_n^0$.
%(as defined in \cref{def:buildingBlocks}). 

%Since $m_1$ divides $p$, and $p$ is even, we require $F$ to be consistent with $\IP_n^0$. Using \cref{cor:consistencyOfBuildingBlocks} and the fact that each of $\matching$, $\timp$, $\nand$ is consistent with $\IP_n^0$, we indeed have that $F$ is consistent with $\IP_n^0$.

%Using \cref{fact:spectraOfDisjointCNFs} and the spectra of these building blocks from \cref{def:buildingBlocks}, we see that 
The spectrum of $F$ is 
$P(x, y, z) = (x^2 + 2y^2 + z^2)^A (x^2 + (y + z)^2)^B ((2y+z)^2)^C$.
We now apply \cref{lem:coeffBoundIntegerCase}
%and use the fact that the coefficient of $x^py^qz^r$ in $P$ determines $\abs{\sol(F)\cap S(p, q, r)}$, to infer that 
to get
\[
\abs{\sol(F)\cap S(p, q, r)} \ge n^{-K_1}\cdot \frac{5^n}{2^{2p+q}}
\]
where $K_1$ is a constant.
Since $\abs{S(p, q, r)} = \binom{n}{p}\binom{n-p}{r}\cdot 2^{n - p - r}$.
Using these, we get
\begin{align*}
\frac{\abs{S(p, q, r)}}{\abs{\sol(F)\cap S(p, q, r)}} 
& \le \binom{n}{p}\binom{n-p}{r}\cdot 2^{n - p - r}\cdot n^{-K_1}\cdot \frac{2^{2p+q}}{5^n}\\
& = n^{-K_1}\cdot \binom{n}{p, q, r}\cdot 4^{-r}\cdot(4/5)^n
\end{align*}
where the last equality follows from the fact that $p + q + r = n$.
To prove 
$\frac{S(p, q, r)}{\abs{\sol(F)\cap S(p, q, r)}} \le n^{-K_0}\cdot (9 / 5)^n$, it suffices to show that 
$\binom{n}{p, q, r}\cdot 4^{-r}\le (9 / 4)^n$ where $K_2$ is a constant. 
We have
\begin{align*}
\binom{n}{p, q, r}\cdot 4^{-r}   
& = \binom{n}{p, q, r} 1^p\cdot 1^q\cdot 4^{-r}\\
& \le \sum_{i, j, k\in \N: i + j + k = n} \binom{n}{i, j, k}\cdot 1^i \cdot 1^j\cdot 4^{-k}\\
& = (9 / 4)^n
\end{align*}
where the last equality follows from the binomial theorem.
Hence, our construction $F$ indeed has the desired properties.
\end{proof}

\subsection{2-CNFs for orbits in region 2}\label{subsubsec:2CNFOrbitRegion2}
In this subsection, we will prove \cref{lem:ip-orbit-cnf-inf-region} for orbits in $\cR_2$. In the proof, we use the following general claim regarding binomial coefficients which will  be proved at the end of the subsection.
\begin{claim}
\label{claim:Region2-UsefulBinomialInequality}
For all $m, k\in \N$ such that $m\ge k$ and $m, k$ are even, we have that
\[
    \binom{m}{k} \le (m/2+1)^2\binom{m/2}{k/2}^2
\]
\end{claim}
%We prove this claim later towards the end of this subsection. Let's see how using it, our proof follows.

\begin{proof}[Proof of \cref{lem:ip-orbit-cnf-inf-region} for region $2$]
Let $m_2 = 4$ and consider any $(p, q, r)\in \cR_2$ such that $m_2$ divides each of $p$, $q$, and $r$ and hence $n$.
%We will construct $2$-CNF $F$ that is consistent with $\IP_n^{p\pmod 2} = \IP_n^0$ (since $p$ is even) and such that $\frac{S(p, q, r)}{\abs{\sol(F)\cap S(p, q, r}} \le n^{K_0} (9 / 5)^n$ where $K_0$ is a  constant.

Let $F$ be the $2$-CNF obtained by taking the disjoint conjunction of $n/2$ copies of $\matching$.
Since $\matching$ is consistent with $\IP_n^0$, $F$ is also consistent with $\IP_n^0$ as desired.
%\cref{cor:consistencyOfBuildingBlocks} to infer .
The spectrum of $F$ is given by $P(x, y, z) = (x^2 + 2y^2 + z^2)^{n/2}$ and that $\abs{\sol(F)\cap S(p, q, r)}$ equals the coefficient of the monomial $x^py^qz^r$ in $P(x, y, z)$.
%We now show the desired upper bound on $\frac{S(p, q, r)}{\abs{\sol(F)\cap S(p, q, r}}$.
%We apply \cref{fact:spectraOfDisjointCNFs} to obtain that 
Examining $P$, we easily deduce that this coefficient equals $\binom{n/2}{p/2}\binom{n/2 - p/2}{q/2} 2^{q/2}$.
Since $S(p,q,r)= \binom{n}{p}\binom{n-p}{r}2^q$, we get 
\[
\frac{S(p, q, r)}{\abs{\sol(F)\cap S(p, q, r}} 
= \frac{\binom{n}{p}\binom{n-p}{r}2^q}{\binom{n/2}{p/2}\binom{(n-p)/2}{q/2}2^{q/2}} 
= \frac{\binom{n}{p}}{\binom{n/2}{p/2}}\cdot\frac{\binom{n-p}{q}}{\binom{(n-p)/2}{q/2}}\cdot 2^{q/2}
\]
where the last equality follows using the fact that $p + q + r = n$.
We apply \cref{claim:Region2-UsefulBinomialInequality} to the last expression and obtain:
\begin{align*}
\frac{S(p, q, r)}{\abs{\sol(F)\cap S(p, q, r}} 
& \le n^4 \cdot \binom{n/2}{p/2}\cdot \binom{(n - p)/2}{q/2}\cdot 2^{q/2}\\
& \le n^4 \cdot \binom{n/2}{p/2}\cdot \sum_{i=0}^{(n-p)/2}\binom{(n - p)/2}{i}\cdot 2^{i}\\
& = n^4 \cdot \binom{n/2}{p/2}\cdot 3^{(n - p) / 2} & \textrm{(by binomial theorem)}
\end{align*}
Since we are in $\cR_2$, we have $p \ge p_0 = 16n / 50\ge (n/2) / 2$,  which implies that the term $\binom{n/2}{p/2}$ decreases as $p$ increases.
Also the term $3^{(n - p) / 2}$ decreases as $p$ increases.
Hence, $n^4 \cdot \binom{n/2}{p/2}\cdot 3^{(n - p) / 2} \le n^4 \cdot \binom{n/2}{p_0/2}\cdot 3^{(n - p_0) / 2}$.
Therefore, we obtain
\[
\frac{S(p, q, r)}{\abs{\sol(F)\cap S(p, q, r}} 
\le n^4 \cdot \binom{n/2}{p_0/2}\cdot 3^{(n - p_0) / 2}
\]
We apply \cref{thm:binomialApprox} to the last expression to get the desired upper bound on $\frac{S(p, q, r)}{\abs{\sol(F)\cap S(p, q, r}}$:
\[
\frac{S(p, q, r)}{\abs{\sol(F)\cap S(p, q, r}} 
\le n^4 \cdot 2^{(n/2)H(p_0 / n)}\cdot 3^{(n - p_0) / 2}
= n^4 \cdot \left(2^{H(16/25) / 2}\cdot 3^{9/50}\right)^n
< n^4 \cdot \left(1.7\right)^n
< n^K\cdot (9 / 5)^n
\]
where $H$ is the binary entropy function and $K = 4$.
%Hence, we obtain the desired upper bound on $\frac{S(p, q, r)}{\abs{\sol(F)\cap S(p, q, r}}$, showing $F$ indeed satisfies all the properties we claimed.
\end{proof}

We now prove our remaining claim, an inequality involving binomial coefficients.
\begin{proof}[Proof of \cref{claim:Region2-UsefulBinomialInequality}]
This will follow by couple of applications of \cref{thm:binomialApprox}.
We compute:
\begin{align*}
\binom{m}{k} 
& \le 2^{m H(k / m)} & \textrm{(applying \cref{thm:binomialApprox})}\\
& = (m/2 + 1)^2 \left(\frac{1}{(m/2+1)^2}\cdot 2^{(m/2) H((k/2) / (m/2))}\right)^2\\
& \le (m/2 + 1)^2 \binom{m/2}{k/2}^2 & \textrm{(applying \cref{thm:binomialApprox})}
\end{align*}
as desired.
%(here $H$ is the binary entropy function).
\end{proof}

\subsection{Constructing 2-CNFs for orbits in region 3}\label{subsubsec:2CNFRegion3}

We here prove \cref{lem:ip-orbit-cnf-inf-region} for region $3$. We need the following lemma regarding the coefficients of certain generating functions which will be  proved in \cref{subsec:R3CoeffExtract}.

\begin{lemma}
\label{lem:Region3CoeffBoundIntegerCase}
There exists a  constant $K$ such that the following holds.
Let $B_1, C_1, B_2, C_2, p, q, r, n$ be nonnegative integers such that
$n\ge K$, $p \le 2\cdot \min(B_1, B_2)$, $n = p + q + r, r\le n-8, B_1 = 0.34 n, C_1 = 0.16 n, B_2 = 0.465 n, C_2 = 0.035 n$.
Then, for $i\in [2]$, the coefficient of the monomial $x^p y^q z^r$ in $P_i(x, y, z) = (x^2 + (y+z)^2)^{B_i} ((2y+z)^2)^{C_i}$ is at least $\frac{1}{n^K}$ $\frac{(2v_i+1)^{2C_i}(u_i + (v_i + 1)^2)^{B_i}}{u_i^{p/2}v_i^q}$ where $v_i = \frac{-\beta_i + \sqrt{\beta_i^2 + 8rq}}{4r}, u_i = \frac{p(v_i+1)^2}{2B_i-p}$, and $\beta_i = 4C_i + (2B_i - p) - 3q$.
\end{lemma}

%We will prove this lemma in \cref{subsec:R3CoeffExtract} and for now just assume it holds.  
We also need the following lemma regarding the solution of a bounded optimization problem which is solved using
an optimization solver:
\begin{lemma}\label{lem:Region3BoundedOptimizationProblem}
Let $\hatp, \hatq, \hatr$ be arbitrary non-negative reals satisfying the following constraints:
\begin{itemize}
\item 
$\hatp + \hatq + \hatr = 1$.
\item 
$\frac{1}{2} - \frac{25}{32}\hatp \ge 0$.
\item 
$\frac{5}{4} - \frac{25}{32}\hatp - \frac{25}{4}\hatr \le 0$.
\item 
$\frac{1}{4} - \hatp \le 0$.
\end{itemize}
Let $\hatB_1 = 0.34, \hatC_1 = 0.16, \hatB_2 = 0.465, \hatC_2 = 0.035$.
For $i\in [2]$, let
$\hat{\beta}_i = 4\hatC_i + (2\hatB_i - \hatp) - 3\hatq, v_i = 
\frac{-\hat{\beta}_i + \sqrt{\hat{\beta}_i^2 + 8\hatr \hatq}}{4\hatr}$.
Finally, let
\begin{align*}
T_i = & -\hatp\log(\hatp) -\hatq\log(\hatq) - \hatr\log(\hatr) + \hatq +\\
& \hatq\log(v_i) - 2\hatC_i\log(2v_i + 1) - (2\hatB_i - \hatp)\log(v_i + 1) + \\
& 0.5 (2\hatB_i - \hatp)\log(2\hatB_i-\hatp) + 0.5 \hatp\log(\hatp) - \hatB_i\log(2\hatB_i) 
\end{align*}
Then, $\min(T_1, T_2) \le 0.841$.
\end{lemma}

\begin{proof}
% [Proof of \cref{lem:Region3BoundedOptimization}]
This as an optimization problem where we maximize $\min(T_1, T_2)$ over all $\hatp, \hatq, \hatr$ subject to the constraints. In particular, we substitute $\hatq = 1 - \hatp - \hatr$ and then maximize over all values of $\hatp, \hatr$ subject to the constraints.
To solve this optimization problem, we use IbexOpt, a well known optimization tool that uses numeric interval arithmetic based library IBEX in C++ to solve global optimization problems \cite{Ibex2015}. 
We provide our code for this at 
% \url{https://anonymous.4open.science/r/Inner-Product-671E/region3.bch}.
% \begin{comment}
\url{https://github.com/mjguru/Inner-Product/blob/main/region3.bch}. 
% \end{comment}
We note that the objective function we use in the code uses upper and lower bounds when evaluating logarithms to avoid numerical stability issues. These bounds only increase our objective function and so our upper bound still holds.

\end{proof}

We are now ready to prove \cref{lem:ip-orbit-cnf-inf-region} for region $3$.

\begin{proof}[Proof of \cref{lem:ip-orbit-cnf-inf-region} for region 3]
Let $m_3 = 2000$.
Consider any $(p, q, r)\in \cR_3$ such that $m$ divides each of $p, q, r$ and hence also $n = p + q + r$.
Since $p$ is even, we will construct 2-CNF $F$ that is consistent with $\IP_n^0$ such that $\frac{\abs{S(p, q, r)}}{\abs{\sol(F) \cap S(p, q, r)}} \le n^{K_0}\cdot (9 / 5)^n$ for some constant $K_0$.

When $r = n$, we have $\abs{S(p, q, r) = 1}$ and in that case, we trivially construct a consistent $2$-CNF covering the solution.
Otherwise, since $m_3$ divides $r$ and $r\le n$, it must be that $r\le n - m\le n - 8$ and we use this fact when we apply \cref{lem:Region3CoeffBoundIntegerCase}.

Let $B_1 = 0.34 n, C_1 = 0.16 n, B_2 = 0.465 n, C_2 = 0.035 n$. We easily see that each of them is a positive even integer. For $i\in [2]$, let $F_i$ be the $2$-CNF obtained by a disjoint conjunction of $B_i$ copies of $\timp$ and $2C_i$ copies of $\nand$.
Since $B_i$ and $C_i$ are even, we infer that $F_i$ is consistent with $\IP_n^0$.
We will let $F$ equal one of $F_1, F_2$, whichever one attains larger value of $\abs{\sol(F_i) \cap S(p, q, r)}$.

The spectrum of $F_i$ is $P_i = (x^2 + (y+z)^2)^{B_i}((2y+z)^2)^{C_i}$.
The coefficient of $x^py^qz^r$ in $P_i$ equals $\abs{\sol(F_i) \cap S(p, q, r)}$.

Recall that $\abs{S(p, q, r)} = \binom{n}{p}\binom{n-p}{r}\cdot 2^q$.
We apply \cref{thm:binomialApprox} to get 

\begin{align*}
\abs{S(p, q, r)}\le 2^{p\log(p / n) + q\log(q / n) + r\log(r / n) + q}.
\end{align*}

Using \cref{lem:Region3CoeffBoundIntegerCase} for each $P_i$, we infer that 
\begin{align*}
\abs{\sol(F_i) \cap S(p, q, r)} \ge n^{-K_1}\cdot \frac{(2v_i+1)^{2C_i}(u_i + (v_i + 1)^2)^{B_i}}{(u_i)^{p/2}v_i^q}
\end{align*}
where $v_i = \frac{-\beta_i + \sqrt{\beta_i^2 + 8rq}}{4r}, u_i = \frac{p(v_i+1)^2}{2B_i-p}$, and $\beta_i = 4C_i + (2B_i - p) - 3q$.
Combing the last two inequalities, we get 
\[
    \frac{\abs{S(p, q, r)}}{\abs{\sol(F_i) \cap S(p, q, r)}} \le \frac{2^{p\log(p / n) + q\log(q / n) + r\log(r / n) + q} (u_i)^{p/2}(v_i)^q}{n^{-K_1}(2v_i+1)^{2C_i}(u_i + (v_i + 1)^2)^{B_i}}.
\]
Let $\sigma_i = \frac{\abs{S(p, q, r)}}{\abs{\sol(F_i) \cap S(p, q, r)}}$ and let $\hatp = \frac{p}{n}, \hatq = \frac{q}{n}, \hatr = \frac{r}{n}, \hatB_i = \frac{B_i}{n}, \hatC_i = \frac{C_i}{n}$.
Then we rewrite: 
\begin{align*}
\frac{\log(\sigma_i)}{n} - K_1\log(n) 
\le & -\hatp\log(\hatp) - \hatq\log(\hatq) - \hatr\log(\hatr) + \hatq +\\
& 0.5 \hatp \log(u_i) + \hatq\log(v_i) - 2\hatC_i\log(2v_i+1) - 2\hatB_i\log(u_i + (v_i +  1)^2).
\end{align*}
We substitute $u_i = \frac{p(v_i+1)^2}{2B_i - p}$ and simplify to obtain:
\begin{align*}
\frac{\log(\sigma_i)}{n} - K_1\log(n) 
\le & \hatp\log(\hatp) + \hatq\log(\hatq) + \hatr\log(\hatr) + \hatq +\\
& \hatq\log(v_i) - 2\hatC_i\log(2v_i + 1) - (2\hatB_i - \hatp)\log(v_i + 1) + \\
& 0.5 (2\hatB_i - \hatp)\log(2\hatB_i-\hatp) + 0.5 \hatp\log(\hatp) - \hatB_i\log(2\hatB_i) 
\end{align*}

We now apply \cref{lem:Region3BoundedOptimizationProblem}, verifying that we satisfy all the constraints, to infer that 
\begin{align*}
\min\left(\frac{\log(\sigma_1)}{n} - K_1\log(n), \frac{\log(\sigma_2)}{n} - K_1\log(n)\right)    
\le 0.8398 & \tag{$2$} \label{eq:Region3ComputerAssistedBound} 
\end{align*}

By definition of $F$, we have that
$\frac{\abs{S(p, q, r)}}{\abs{\sol(F) \cap S(p, q, r)}} = \min\left(\sigma_1, \sigma_2\right)$.
Hence, we compute that
\begin{align*}
\frac{\abs{S(p, q, r)}}{\abs{\sol(F) \cap S(p, q, r)}}
& = \min\left(\sigma_1, \sigma_2\right)\\
& = 2^{n \cdot \min\left(\frac{\log(\sigma_1)}{n}, \frac{\log(\sigma_1)}{n}\right)}\\
& \le n^{K_1}\cdot 2^{n\cdot 0.8398} & \textrm{(using \cref{eq:Region3ComputerAssistedBound})}\\
& \le n^{K_1}\cdot (9 / 5)^n
\end{align*}
as desired.
\end{proof}

\subsection{Constructing 2-CNFs for orbits in region 4}\label{subsubsec:2CNFRegion4}

In this subsection, We prove \cref{lem:ip-orbit-cnf-inf-region} for region $4$. We need the following lemma regarding the coefficients of certain generating functions which itself will be proved in
\cref{subsec:R4CoeffExtract}

\begin{lemma}
\label{lem:R4CoeffBoundIntegerCase}
There exists a  constant $K$ such that the following holds.
Let $A_1, C_1, A_2, C_2, p, q, r, n$ be nonnegative integers such that
$n\ge K_0$, $p \le 2\cdot \min(A_1, A_2)$, $n = p + q + r, 0 < r\le n-8, A_1 = 0.34 n, C_1 = 0.16 n, A_2 = 0.355 n, C_2 = 0.145 n$.
Also, for $i\in [2]$, define the degree $3$ real polynomial $Q_i(x) = (4r) x^3 + (4A_i - 2p - 2q)x^2 + (4C_i - 2q) x - q$.
Then, for $i\in [2]$, the following holds:
(1) $Q_i(x)$ has a unique non-negative real root.
(2) The coefficient of the monomial $x^p y^q z^r$ in $P_i(x, y, z) = (x^2 + 2y^2 + z^2)^{A_i} ((2y+z)^2)^{C_i}$ is at least $\frac{1}{n^K}$ $\frac{(2v_i+1)^{2C_i}(u_i + 2v_i^2 + 1)^{A_i}}{u_i^{p/2}v_i^q}$ where $v_i$ is the unique non-negative root of the cubic polynomial $Q_i(x)$, and $u_i = \frac{p(2v_i^2 + 1)}{2A_i - p}$.
\end{lemma}

%We will prove this in \cref{subsec:R4CoeffExtract} and for now just assume it holds.  
We also need the solution of a bounded optimization problem which is obtained using an optimization solver.

\begin{lemma}\label{lem:Region4BoundedOptimizationProblem}
Let $\hatp, \hatq, \hatr$ be arbitrary non-negative reals satisfying the following constraints:
\begin{itemize}
\item 
$\hatp + \hatq + \hatr = 1$.
\item 
$\frac{1}{2} - \frac{25}{32}\hatp \ge 0$.
\item 
$-\frac{5}{4} + \frac{25}{16}\hatp + \frac{25}{4}\hatr \le 0$.
\item 
$-\frac{1}{20} + \hatr \ge 0$.

\item 
$\frac{1}{4} - \hatp \le 0$.
\end{itemize}
Let $\hatA_1 = 0.34, \hatC_1 = 0.16, \hatA_2 = 0.355, \hatC_2 = 0.145$.
For $i\in [2]$, let
$v_i$ be the unique non-negative root of the cubic polynomial $P_i(x) = (4\hatr) x^3 + (4\hatA_i - 2\hatp - 2\hatq)x^2 + (4\hatC_i - 2\hatq) x - \hatq$, and let $u_i = \frac{p(2v_i^2 + 1)}{2\hatA_i - \hatp}$
Finally, let
\begin{align*}
T_i = 
& -\hatp\log(\hatp) -\hatq\log(\hatq) -\hatr\log(\hatr) + \hatq +\\
& -2\hatC_i \log(2\hatv_i + 1) -\hatA_i\log(\hatu_i + 2\hatv_i^2 + 1) +\\
& (\hatp/2)\log(\hatu_i) + \hatq\log(\hatv_i)
\end{align*}
Then, $\min(T_1, T_2) \le 0.845$.
\end{lemma}

\begin{proof}
% [Proof of \cref{lem:Region4BoundedOptimizationProblem}]
We will first assert the existence and uniqueness of the nonnegative real root of the polynomial $P_i(x)$ which follows from \cref{lem:R4CoeffBoundIntegerCase}.

We now describe how we obtain the upper bound on $\min(T_1, T_2)$.
This as an optimization problem where we maximize $\min(T_1, T_2)$ over all $\hatp, \hatq, \hatr$ subject to the constraints. In particular, we substitute $\hatq = 1 - \hatp - \hatr$ and then maximize over all values of $\hatp, \hatr$ subject to the constraints.
To solve this optimization problem, we use IbexOpt, a well known optimization tool that uses numeric interval arithmetic based library IBEX in C++ to solve global optimization problems \cite{Ibex2015}. 
We provide our code for this at 
% \url{https://anonymous.4open.science/r/Inner-Product-671E/region4.bch}.
% \begin{comment}
\url{https://github.com/mjguru/Inner-Product/blob/main/region4.bch}. 
% \end{comment}
We note that the objective function we use in the code uses upper and lower bounds when evaluating logarithms to avoid numerical stability issues. These bounds only increase our objective function and so our upper bound still holds. 

\end{proof}

We are now ready to prove \cref{lem:ip-orbit-cnf-inf-region} for region 4.

\begin{proof}[Proof of \cref{lem:ip-orbit-cnf-inf-region} for region 4]
We let $m_4 = 2000$.
Consider any $(p, q, r)\in \cR_4$ such that $m$ divides each of $p, q, r$ and hence also $n = p + q + r$.
Since $p$ is even, we will construct a 2-CNF $F$ that is consistent with $\IP_n^0$ such that $\frac{\abs{S(p, q, r)}}{\abs{\sol(F) \cap S(p, q, r)}} \le n^{K_0}\cdot (9 / 5)^n$ where $K_0$ is a  constant.

Let $A_1 = 0.34 n, C_1 = 0.16 n, A_2 = 0.355 n, C_2 = 0.145 n$. Since $2000$ divides $n$, each of these is a positive even integer. 
For $i\in [2]$, let $F_i$ be the $2$-CNF obtained by a disjoint conjunction of $A_i$ copies of $\matching$ and $2C_i$ copies of $\nand$.
Using \cref{cor:consistencyOfBuildingBlocks} and the fact that each of $A_i$ and $C_i$ are even integers, we infer that $F_i$ is indeed consistent with $\IP_n^0$.
Let $F$ equal one of $F_1, F_2$, whichever one attains larger value of $\abs{\sol(F_i) \cap S(p, q, r)}$.

The spectrum of $F_i$ is $P_i = (x^2 + 2y^2 + z^2)^{A_i}((2y+z)^2)^{C_i}$.
The coefficient of $x^py^qz^r$ in $P_i$ equals $\abs{\sol(F_i) \cap S(p, q, r)}$.

We know that $\abs{S(p, q, r)} = \binom{n}{p}\binom{n-p}{r}\cdot 2^q$.
We apply \cref{thm:binomialApprox} and simplify using the fact  $p + q + r = n$ to infer that
$\abs{S(p, q, r)}\le 2^{p\log(p / n) + q\log(q / n) + r\log(r / n) + q}$. 

Using \cref{lem:R4CoeffBoundIntegerCase} for each $P_i$, we infer that 
\begin{align*}
\abs{\sol(F_i) \cap S(p, q, r)} \ge n^{-K_1}\cdot \frac{(2v_i+1)^{2C_i}(u_i + 2v_i^2 + 1)^{A_i}}{(u_i)^{p/2}v_i^q}
\end{align*}
where $v_i$ is the unique non-negative root of the polynomial $(4r)x^3 + (4A_i - 2p-2q)x^2 + (4C_i - 2q)x - q$,  $u_i = \frac{p(2v_i^2 + 1)}{2A_i - p}$ and $K_1$ a constant.
We now have
\[
\frac{\abs{S(p, q, r)}}{\abs{\sol(F_i) \cap S(p, q, r)}} 
\le \frac{2^{p\log(p / n) + q\log(q / n) + r\log(r / n) + q} (u_i)^{p/2}(v_i)^q}{n^{-K_1}(2v_i+1)^{2C_i}(u_i + 2v_i^2 + 1)^{A_i}}.
\]
Let $\sigma_i = \frac{\abs{S(p, q, r)}}{\abs{\sol(F_i) \cap S(p, q, r)}}$ and let $\hatp = \frac{p}{n}, \hatq = \frac{q}{n}, \hatr = \frac{r}{n}, \hatB_i = \frac{B_i}{n}, \hatC_i = \frac{C_i}{n}$.
We  write the inequality as 
\begin{align*}
\frac{\log(\sigma_i)}{n} - K_1\log(n) 
\le & -\hatp\log(\hatp) -\hatq\log(\hatq) -\hatr\log(\hatr) + \hatq +\\
& 2\hatC_i\log(2v_i + 1) - \hatA_i \log(u_i + 2v_i^2 + 1) + \\
& (\hatp/2)\log(u_i) + \hatq\log(v_i)
\end{align*}

After verifying that we satisfy all the constraints of the lemma (since $(p,q,r)\in \cR_5$, we must have that $r\le 0.9n\le n-8$), we now apply \cref{lem:Region4BoundedOptimizationProblem},  to infer that 
\begin{align*}
\min\left(\frac{\log(\sigma_1)}{n} - K_1\log(n), \frac{\log(\sigma_2)}{n} - K_1\log(n)\right)    
\le 0.844 & \tag{$2$} \label{eq:Region4ComputerAssistedBound} 
\end{align*}

By definition of $F$, we have that
$\frac{\abs{S(p, q, r)}}{\abs{\sol(F) \cap S(p, q, r)}} = \min\left(\sigma_1, \sigma_2\right)$.
Hence, we conclude that
\begin{align*}
\frac{\abs{S(p, q, r)}}{\abs{\sol(F) \cap S(p, q, r)}}
& = \min\left(\sigma_1, \sigma_2\right)\\
& = 2^{n \cdot \min\left(\frac{\log(\sigma_1)}{n}, \frac{\log(\sigma_1)}{n}\right)}\\
& \le n^{K_1}\cdot 2^{n\cdot 0.844} & \textrm{(using \cref{eq:Region4ComputerAssistedBound})}\\
& \le n^{K_1}\cdot (9 / 5)^n
\end{align*}
as desired.
\end{proof}

\subsection{Constructing 2-CNFs for orbits in region 5}\label{subsubsec:2CNFRegion5}

In this subsection, we prove \cref{lem:ip-orbit-cnf-inf-region} for orbits in $\cR_5$.
We need the following result which we prove towards the end of this subsection:

\begin{claim}\label{claim:Region5HelperBound}
Let $g: (0, 0.64] \to \R$  be defined as $g(y) = H(y) + \eps_R H(\eps_R) + a - \frac{y}{2} - a H(y / 2a)$ where $\eps_R = \frac{1}{20}, a = 0.355$ and $H$ is the binary entropy function.
Then, $\max_{y\in [0, 0.64]} g(y) \le 0.828$.
\end{claim}

We first show how \cref{lem:ip-orbit-cnf-inf-region} follows.
\begin{proof}[Proof of \cref{lem:ip-orbit-cnf-inf-region} for region 5]
Let $m_5 = 2000$ and let $(p, q, r)\in \cR_5$ be an arbitrary orbit such that $m$ divides each of $p$, $q$, and $r$ and hence $n$.
Since $p$ is even, it suffices to construct a $2$-CNF $F$ that is consistent with $\IP_n^0$ and  $\frac{\abs{S(p, q, r)}}{\abs{\sol(F)\cap S(p, q, r)}} \le n^{K_0}(9/5)^n$ where $K_0$ is a  constant.

Let $A = 0.355 n, C = 0.145 n$.
Let $F$ be the $2$-CNF obtained by disjoint conjunctions of  $A$ copies of $\matching$ and $2C$ copies of $\nand$ (as defined in \cref{def:buildingBlocks}).
Since both these building blocks are consistent with $\IP_n^0$, we have that $F$ is consistent with $\IP_n^0$ as desired.

We will now show that $\frac{\abs{S(p, q, r)}}{\abs{\sol(F)\cap S(p, q, r)}} \le n^{K_0}(9/5)^n$.
The spectrum of $F$ is given by $P(x, y, z) = (x^2 + 2y^2 + z^2)^A (2y+z)^{2C}$ and $\abs{\sol(F)\cap S(p, q, r)}$ equals the coefficient of $x^py^qz^r$ in $P(x, y, z)$.
We lower bound this coefficient as follows: Choose $r$ out of $2C$ terms of $(2y+z)$ to equal $z$ and remaining to equal $2y$; then choose $p/2$ terms out of $A_i$ terms to be $x^2$ and remaining terms to equal $2y^2$
to obtain the  monomial $2^{A - p/2 + 2C - r}\cdot x^py^qz^r$. Since there are $\binom{2C}{r}\cdot \binom{A}{p/2} \ge \binom{A}{p/2}$ of obtaining such a monomial, we lower bound its coefficient by $\binom{A}{p/2}\cdot 2^{A + 2C - p/2 - r}$.
Since 
$p \le 0.64$ and $r\le 10^{-5}$, we have $\binom{n-p}{r}\le \binom{n}{r}$.
Since $\abs{S(p, q, r)} = \binom{n}{p}\binom{n-p}{r}\cdot 2^{q}$, 
we get
\begin{align*}
\frac{\abs{S(p, q, r)}}{\abs{\sol(F) \cap S(p, q, r)}}    
& \le \frac{\binom{n}{p}\binom{n}{r} 2^{q}}{\binom{A}{p/2} 2^{A + 2C - p/2 - r}}\\
& = \frac{\binom{n}{p}\binom{n}{r} 2^{A - p/2}}{\binom{A}{p/2}} 
\end{align*}
where in the last line we used the fact $n = 2A + 2C = p + q + r$.

We now repeatedly apply \cref{thm:binomialApprox} to the last expression to get the following bound:
\[
\frac{\abs{S(p, q, r)}}{\abs{\sol(F) \cap S(p, q, r)}}    
\le (A+1)\cdot 2^{n\cdot H(p/n) + r\cdot H(r/n) + A - p/2 - A\cdot H(p/(2A))}
\]
where $H$ is the binary entropy function.
Using the fact that $A+1\le n$ and rearranging and taking logarithm of both sides, we have  
\[
    \frac{\log\left(\frac{\abs{S(p, q, r)}}{\abs{\sol(F) \cap S(p, q, r)}}\right)}{n} - \log(n) \le H(\hatp) + \hatr\cdot H(\hatr) + \hatA - \hatp/2 - \hatA\cdot H(\hatp/(2\hatA))
\]
where $\hatA = \frac{A}{n}, \hatp = \frac{p}{n}, \hatr = \frac{r}{n}$.
Since $0\le \hatr \le \frac{1}{20}$ and $\hatr \cdot H(\hatr)$ is an increasing function in that range, we infer 
\begin{align*}
\frac{\log\left(\frac{\abs{S(p, q, r)}}{\abs{\sol(F) \cap S(p, q, r)}}\right)}{n} - \log(n) 
& \le H(\hatp) + \frac{1}{20}\cdot H\left(\frac{1}{20}\right) + \hatA - \hatp/2 - \hatA\cdot H(\hatp/(2\hatA))
\end{align*}
Let $g(y) = H(y) + \frac{1}{20}H\left(\frac{1}{20}\right) + \hatA - y/2 - \hatA\cdot H(y / 2\hatA)$ where $\hatA = 0.355$.
Then, using \cref{claim:Region5HelperBound}, we get that $g(\hatp)\le 0.828$ and 
\[
\frac{\log\left(\frac{\abs{S(p, q, r)}}{\abs{\sol(F) \cap S(p, q, r)}}\right)}{n} - \log(n) \le 0.828
\]
Rearranging and considering exponents, we conclude 
\[
\frac{\abs{S(p, q, r)}}{\abs{\sol(F) \cap S(p, q, r)}}
\le n\cdot 2^{0.828 n} \le n^1\cdot (9/5)^n
\]
as desired.
\end{proof}

We now prove the claim regarding the real function $g$:
\begin{proof}[Proof of \cref{claim:Region5HelperBound}]
We will show that for all $y \in (0, 6.4]$, $g'(y) > 0$ which lets us conclude that $\max_{y\in [0, 0.64]} g(y) = g(0.64) \le 0.828$ as desired.

Recall that the derivative of the binary entropy function $H$ is $H'(x) = \log((1-x) / x)$.
Using this, we compute that
\begin{align*}
g'(y)
& = \log((1 - y) / y) - \frac{1}{2} - \frac{1}{2}\log((2a - y) / y)\\
& = \log\left(\frac{1-y}{\sqrt{2y(2a-y)}}\right)
\end{align*}
Therefore, to show that $g'(y) > 0$ for all $y\in (0, 0.64]$, it suffices to show that for $y$ in this domain, $1-y > \sqrt{2y(2a-y)}$.
Since both sides of this inequality are positive, this is equivalent to showing that 
$(1 - y)^2 > 2y(2a-y)$. 
This is equivalent to showing $3y^2 - y(4a + 2) +  1 > 0$.
Let $h(y) = 3y^2 - y(4a + 2) +  1$.
We will show that $h(y) > 0$ for all $y$.
The discriminant of this quadratic $h(y)$ is $(4a+2)^2 - 12 = (3.42)^2 - 12 < 0$. This means $h(y)$ has no real roots. Since $h(0) = 1 > 0$, we conclude that $h(y) > 0$  for all $y$ as desired.
\end{proof}

\subsection{Constructing 2-CNFs for orbits in region 6}\label{subsubsec:2CNFRegion6}

%In this subsection, we will prove \cref{lem:ip-orbit-cnf-inf-region} for triplets in $\cR_6$.

\begin{proof}[Proof of \cref{lem:ip-orbit-cnf-inf-region} for region 6]
Let $m_6 = 2000$ and let $(p, q, r)\in \cR_6$ be an arbitrary orbit such that $m$ divides each of $p$, $q$ and $r$ and hence $n$. 
Since $p$ is even, it suffices to show that we can construct a $2$-CNF $F$ that is consistent with $\IP_n^0$ such that $\frac{\abs{S(p, q, r)}}{\abs{\sol(F)\cap S(p, q, r)}} \le n^{K_0}(9/5)^n$ for some constant $K_0$.

Let $A = p/2, C = (n - p) / 2$.
Let $F$ be the $2$-CNF obtained by a disjoint conjunction of  $A$ copies of $\matching$ and $2C$ copies of $\nand$.
Since both these building blocks are consistent with $\IP_n^0$, we have that $F$ is consistent with $\IP_n^0$ as desired.

We will now show that $\frac{\abs{S(p, q, r)}}{\abs{\sol(F)\cap S(p, q, r)}} \le n^{K_0}(9/5)^n$.
The spectrum of $F$ is given by $P(x, y, z) = (x^2 + 2y^2 + z^2)^A (2y+z)^{2C}$ and $\abs{\sol(F)\cap S(p, q, r)}$ equals the coefficient of $x^py^qz^r$ in $P(x, y, z)$.
Since $A = p/2$, this coefficient exactly equals
$\binom{2C}{q}\cdot 2^q = \binom{n-p}{q}\cdot 2^q$
Also
%, using \cref{fact:IPOrbitSize}, 
we have that $\abs{S(p, q, r)} = \binom{n}{p}\binom{n-p}{q}\cdot 2^{q}$.
Therefore,
\[
\frac{\abs{S(p, q, r)}}{\abs{\sol(F)\cap S(p, q, r)}} = \binom{n}{p}.
\]
We apply \cref{thm:binomialApprox} to get that
\[
\frac{\abs{S(p, q, r)}}{\abs{\sol(F)\cap S(p, q, r)}} \le 2^{n\cdot H(p/n)}
\]
Since $0\le p\le \frac{n}{4}$ and $H(\cdot)$, the binary entropy function, is increasing in this range, we obtain
\[
\frac{\abs{S(p, q, r)}}{\abs{\sol(F)\cap S(p, q, r)}} \le 2^{n\cdot H(1/4)} \le 2^{n\cdot (0.82)} \le (9 /5)^n
\]
as desired.
\end{proof}

\section{Coefficient extraction for each region}\label{sec:coeffExtract}

In this section, we obtain asymptotic bounds for coefficients of monomials of particular generating functions. 
%These generating functions and monomials will arise from some region from \cref{subsec:construction-2cnf-orbit}.
To prove these results, we require the following lemma which bounds the coefficients of a power series:

\begin{lemma}
\label{lem:bivariateOptApproxCauchy}
Let $p, q \in \N$ be such that $p^2 + q^2 \ge 4$.
Let $f: \C^2\to \C$ be analytic on a compact neighborhood $\cN$ around $(0, 0)$.
Let $f(u, v) = \sum_{r = (r_1, r_2)\in \N^2} C_r u^{r_1}v^{r_2}$ where for $C_r$ are non-negative.
%\in \R$ and $C_i \ge 0$. 
Let $h: \C^2 \to \C$ be defined as $h(u, v) = \ln(f(u, v)) - p\ln(u) - q\ln(v)$.
Assume there exist non-negative real $u_0$ and $v_0$ such that $\nabla_h(u_0, v_0) = 0$ and that $\cH_h(u_0, v_0)$ is positive definite.
Assume that $\abs{f(u, v)}$ attains its unique global maximum on the torus $\abs{u} = u_0, \abs{v} = v_0$ at $(u_0, v_0)$.
Then,  
\[
    C_{p, q} = \frac{f(u_0, v_0)}{2\pi u_0^{p+1}v_0^{q+1} \sqrt{\det(\cH_{h}(u_0, v_0))}}\left(1 + O\left(\frac{1}{\sqrt{p^2 + q^2}}\right)\right).
\]
where $\cH_{h}(u_0, v_0)$ is the Hessian of $h$ evaluated at $u_0, v_0$.
\end{lemma}
This  result seems to be a standard result in analytic combinatorics but since we could not find a ready reference to the formulation in \cref{lem:bivariateOptApproxCauchy}, we will provide a somewhat self-contained proof  in \cref{subsec:bivariateOptApproxCauchy}. For now, we use \cref{lem:bivariateOptApproxCauchy} to obtain bounds on the coefficients of the generating functions that arise from the constructions of $\IP$.
%In \cref{subsec:R1CoeffExtract} these will come from $\cR_1$, in \cref{subsec:R3CoeffExtract}, these will come from $\cR_3$, and in \cref{subsec:R4CoeffExtract}, these will come from $\cR_4$.

\subsection{Bound for Region 1}\label{subsec:R1CoeffExtract}

%In this subsection, we obtain asymptotic bound on coefficient of generating function from $\cR_1$.

\begin{proof}[Proof of \cref{lem:coeffBoundIntegerCase}]
Let $f(u, v) = P(u^{1/2}, v, 1)$.
Since the degree of $x$ is even in $P(x,y,z)$,  
%appears in squared form in each of the three polynomials, it is still the case that $f(u, v)$ is a product of powers of three polynomials.
$2A + 2B + 2C = p + q + r = n$,  and each of the three factor polynomials is homogeneous, we infer that the coefficient of $u^{p/2} v^q$ in $f(u, v)$ equals the coefficient of $x^py^qz^r$ in $P(x, y, z)$.

Let $h: \C^2 \to \C$ as $h(u, v) = \ln(f(u, v)) - (p/2)\ln(u) - q\ln(v)$.
We observe that $\nabla_h(16, 2) = 0$.
We argue  that the Hessian  $\cH_h(16, 2)$ of $h$ at $u=16$ and $v=2$ is positive definite.
To show this, let $\alpha(x, y) = h(\exp(x), \exp(y))$.
We know  the function $\alpha$ is convex since  that $h$ has nonnegative coefficients in its power series (see section 4.5 from \cite{bv2004convex}). We also observe that $\alpha$ is strictly convex and hence $\cH_\alpha(16, 2)$ is positive definite.
%By picking any three non-zero terms of the power series and checking that they don't lie in a line, we see that $\alpha$ will be strictly convex 
Since $\nabla_h(16, 2) = 0$, we apply the chain rule to get that
$\cH_{\alpha}(16, 2) = D\cdot \cH_h(16, 2)\cdot D$ where $D = \begin{bmatrix}16& 0 \\ 0& 2\end{bmatrix}$ is a diagonal matrix.
Since $\cH_{\alpha}(16, 2)$ is positive definite and $D$ is positive definite, we get that $\cH_h(16, 2)$ is positive definite as well.
% Since $\exp$ is monotonic, we infer that there is a unique minima for $h$ as well, and in particular, the point $(16, 2)$ is the unique point (in the domain of non-negative coordinates) where $\nabla_h(16, 2) = 0$.
% Hence, this must also be true in any small neighborhood around $(16, 2)$, making $\cH_h(16, 2)$ positive definite.
We next apply triangle inequality to infer that in the (complex) torus defined by $\abs{u} = 16, \abs{v} = 2$, the unique global maximum occurs at $u = 16$ and $v=2$.
%(we need to use that at $(\pm 16, \pm 2)$, the value is smaller).
Lastly, we claim that at least one of $p$ or $q$ must have value at least $0.01 n$. Indeed, if they both are smaller, then it must be that $r\ge 0.98 n$, implying that $A < 0$. However, that is a contradiction since we assume $A$ is a nonnegative integer.
Let $C_0$ be large enough so we have  $(p/2)^2 + q^2 \ge 4$ for $n\geq C_0$.

We satisfy the conditions to apply \cref{lem:bivariateOptApproxCauchy} with $u_0 = 16, v_0 = 2$ to extract the coefficient $u^{p/2}v^q$.
We make sure that $C_0$ is large enough to conclude that the coefficient $u^{p/2}v^q$
equals
\begin{align*}
\frac{25^{A + B + C}}{C_0 4^p 2^q \poly(p, q)}\left(1 + \frac{1}{\sqrt{p^2 + q^2}}\right)
& \ge \frac{5^n}{C_0 4^p 2^q \poly(n)}\left(1 + \frac{1}{\poly(n)}\right)
\end{align*}
where in the last equality we used the fact that either $p$ or $q$ will be at least $0.01 n$.
%Hence, our desired lower bound on the coefficient indeed holds.
\end{proof}

\subsection{Bound for region 3}\label{subsec:R3CoeffExtract}

In this subsection, we obtain asymptotic bounds on the coefficients of generating functions related to $\cR_3$.

\begin{proof}[Proof of \cref{lem:Region3CoeffBoundIntegerCase}]
For $i\in [2]$, let $f_i(u, v) = P_i(u^{1/2}, v, 1)$. Since $x$ appears in squared form, $f_i$ is still product of three polynomials. Furthermore, since $2B_i + 2C_i = p + q + r = n$, and both the polynomials are homogeneous, we infer that the coefficient of $u^{p/2}v^q$ in $f_i(x, y, z)$ equals the coefficient of $x^p y^q z^r$ in $P_i(x, y, z)$.

Define $h_i: \C^2 \to \C$ by $h(u, v) = \ln(f_i(u, v)) - (p/2)\ln(u) - q\ln(v)$.
We will find non-negative $(u_i, v_i)\in \R^2$ such that $\nabla_{h_i}(u_i, v_i) = 0$.
We compute that $\frac{\partial h_i}{\partial u} = -\frac{p}{2}\cdot \frac{1}{u} + \frac{B_i}{u + (v+1)^2}$.
Setting this to $0$ and rearranging, we infer that
\begin{align*}
u_i = \frac{p (v_i + 1)^2}{2B_i - p}  & \tag{$1$} \label{eq: region 3 value of ui}  
\end{align*}
We next compute that $\frac{\partial h_i}{\partial v} = \frac{4C_i}{2v+1} + \frac{2B_i(v+1)}{u + (v+1)^2} - \frac{q}{v}$.
Setting this to $0$, rearranging, substituting in \cref{eq: region 3 value of ui}, and using the fact that $4C_i + 2(2B_i-p) - 2q = 2r$ (since $2B_i+2C_i = p+q+r = n$), we infer that $v_i$ satisfies the following quadratic equation:
\[
    2r\cdot (v_i)^2 + \beta_i\cdot v_i - q = 0
\]
where $\beta_i = 4C_i + (2B_i - p) - 3q$.
This means $v_i$ equals $\frac{\beta_i \pm \sqrt{\beta_i^2 + 8rq}}{4r}$.
By the inequalities satisfied by $p, q, r$, it must be the case that $rq$ is at least a  constant. 
This makes the value inside the square root comes out to be strictly larger than $\beta_i^2$ and so one root is strictly negative and other is strictly positive.
We reject the negative root since \cref{lem:bivariateOptApproxCauchy} requires the special point $(u_i, v_i)$ to have non-negative real values.

With this value of $u_i, v_i$, we show that 
$\cH_{h_i}(u_i, v_i)$ is positive definite where $\cH_{h_i}$ is the the Hessian of $h_i$. For this, let $\alpha_i(x, y) = h_i(\exp(x), \exp(y))$. Using the fact that $h$ has nonnegative coefficients in its power series, we use the well known result (see section 4.5 from \cite{bv2004convex}) that the function $\alpha_i$ as we defined is convex.
Also by picking any three non-zero terms of the power series and checking that they don't lie in single line, we see that $\alpha_i$ is strictly convex and so $\cH_{\alpha_i}(u_i, v_i)$ is positive definite. Using the fact that $\nabla_{h_i}(u_i, v_i) = 0$ and chain rule, we obtain that $\cH_{\alpha_i}(u_i, v_i) = D\cdot \cH_{h_i}(u_i, v_i)\cdot D$ where $D$ is the diagonal matrix with diagonal entries $u_i$ and $v_i$.
Since $D$ is positive definite, we infer that $\cH_{h_i}(u_i, v_i)$ is indeed positive definite.

We next apply the triangle inequality to infer that in the complex torus with $\abs{u} = u_i, \abs{v} = v_i$, the unique global maximum for $f_i$ occurs at $(u_i, v_i)$ (carefully checking that the values at $(\pm u_i, \pm v_i)$ are smaller).  
Also since $r \le n - 8$, $p + q + r = n$ and $n\ge K_0$ where we will set $K_0$ to be large enough  constant, it must be the case that $(p/2)^2 + q^2 \ge 4$.
Therefore we satisfy all conditions to apply \cref{lem:bivariateOptApproxCauchy} with $(u_0, v_0) = (u_i, v_i)$, we infer that the coefficient of $u^{p/2}v^q$ in $f_i(u, v)$ equals
\[
    \frac{f(u_i, v_i)}{2\pi u_i^{p+1}v_i^{q+1} \sqrt{\cH_{h_i}(u_i, v_i)}} \left(1 + O\left(\frac{1}{\sqrt{p^2 + q^2}}\right)\right).
\]
Since $p^2 + q^2 \ge 4$, and $u_i, v_i\le n$, and that the entries of $\cH_{h_i}$ are bounded by $\poly(u_i, v_i)$, we infer that there exists a  constant $K_1$ such that the coefficient is at least
\[
    n^{-K_1}\cdot \frac{f(u_i, v_i)}{u_i^{p/2} v_i^q}
    = n^{-K_1}\cdot \frac{(u_i + (v_i + 1)^2)^{B_i}(2v_i+1)^{2C_i}}{u_i^{p/2} v_i^q}
\]
\end{proof}

\subsection{Bound for region 4}\label{subsec:R4CoeffExtract}

In this subsection, we obtain asymptotic bounds on the coefficients of generating functions related to $\cR_4$.

\begin{proof}[Proof of \cref{lem:R4CoeffBoundIntegerCase}]
We first prove that $Q_i$ has a unique real nonnegative root.
First, since $Q_i$ has degree $3$, it has at least one real root.
Let $x_0$ be any such real root.
Since $Q_i(x_0) = 0$, we rearrange to infer that $x_0$ must satisfy:
\[
    \frac{4C_ix_0}{2x_0 + 1} + \frac{2x_0^2 (2A_i - p)}{2x_0^2 + 1} = q
\]
We see that the left side at $x = 0$ equals $0$
and as $x\rightarrow +\infty$, it approaches $2C_i + 2A_i - p = q + r$ (since $n = 2A_i + 2C_i = p + q + r$).
Since $q$ and $r$ are nonnegative and right side equals $q$, by intermediate value theorem, there must exist a unique value $y\ge 0$ such that left side equals $q$, proving our result.

We now prove the asymptotic bound.
For $i\in [2]$, let $f_i(u, v) = P_i(u^{1/2}, v, 1)$. Since $x$ appears in squared form, $f_i$ is still product of three polynomials. Furthermore, since $2A_i + 2C_i = p + q + r = n$, and both the polynomials are homogeneous, we infer that the coefficient of $u^{p/2}v^q$ in $f_i(x, y, z)$ equals the coefficient of $x^p y^q z^r$ in $P_i(x, y, z)$.

Define $h_i: \C^2 \to \C$ by $h(u, v) = \ln(f_i(u, v)) - (p/2)\ln(u) - q\ln(v)$.
We will find non-negative $(u_i, v_i)\in \R^2$ such that $\nabla_{h_i}(u_i, v_i) = 0$.
We compute that $\frac{\partial h_i}{\partial u} = \frac{A_i}{u + 2v^2 + 1} - \frac{p}{2u}$.
Setting this to $0$ and rearranging, we infer that
\begin{align*}
u_i = \frac{p (2v_i^2 + 1)}{2A_i - p}  & \tag{$1$} \label{eq: region 4 value of ui}  
\end{align*}
We next compute that $\frac{\partial h_i}{\partial v} = \frac{4C_i}{2v+1} + \frac{4vA_i}{u + 2v^2 + 1} - \frac{q}{v}$.
Setting this to $0$, rearranging, substituting in \cref{eq: region 4 value of ui}, 
and using the fact that $8C_i + 4(2A_i-p) - 4q = 4r$ (since $2A_i+2C_i = p+q+r = n$), 
we infer that $v_i$ satisfies the following cubic equation:
\[
    (4r)\cdot v_i^3 + (4A_i - 2p - 2q)\cdot v_i^2 + (4C_i - 2q)\cdot v_i - q = 0
\]
In other words, $Q_i(v_i) = 0$.
From above, we know that there is a unique nonnegative root of $Q_i$ and we let $v_i$ be the unique nonnegative root.

With this value of $u_i, v_i$, we show that 
$\cH_{h_i}(u_i, v_i)$ is positive definite where $\cH_{h_i}$ is the the Hessian of $h_i$. For this, let $\alpha_i(x, y) = h_i(\exp(x), \exp(y))$. Using the fact that $h$ has nonnegative coefficients in its power series, we use the well known result (see section 4.5 from \cite{bv2004convex}) that the function $\alpha_i$ as we defined is convex.
Also by picking any three non-zero terms of the power series and checking that they don't lie in single line, we see that $\alpha_i$ will be strictly convex and so $\cH_{\alpha_i}(u_i, v_i)$ is positive definite. Using the fact that $\nabla_{h_i}(u_i, v_i) = 0$ and chain rule, we obtain that $\cH_{\alpha_i}(u_i, v_i) = D\cdot \cH_{h_i}(u_i, v_i)\cdot D$ where $D$ is the diagonal matrix with with diagonal entries $u_i$ and $v_i$.
Since $D$ is positive definite, we infer that $\cH_{h_i}(u_i, v_i)$ is indeed positive definite.

We next apply the triangle inequality to infer that in the complex torus with $\abs{u} = u_i, \abs{v} = v_i$, the unique global maximum for $f_i$ occurs at $(u_i, v_i)$ (carefully checking that the values at $(\pm u_i, \pm v_i)$ are smaller).  
Also since $r \le n - 8$, $p + q + r = n$ and $n\ge K_0$ where we will set $K_0$ to be large enough  constant, it must be the case that $(p/2)^2 + q^2 \ge 4$.
Therefore we satisfy all conditions to apply \cref{lem:bivariateOptApproxCauchy} with $(u_0, v_0) = (u_i, v_i)$, we infer that the coefficient of $u^{p/2}v^q$ in $f_i(u, v)$ equals
\[
    \frac{f(u_i, v_i)}{2\pi u_i^{p+1}v_i^{q+1} \sqrt{\cH_{h_i}(u_i, v_i)}} \left(1 + O\left(\frac{1}{\sqrt{p^2 + q^2}}\right)\right).
\]
Since $p^2 + q^2 \ge 4$, and $u_i, v_i\le n$, and that the entries of $\cH_{h_i}$ are bounded by $\poly(u_i, v_i)$, we infer that there exists a  constant $K_1$ such that the coefficient is at least
\[
    n^{-K_1}\cdot \frac{f(u_i, v_i)}{u_i^{p/2} v_i^q}
    = n^{-K_1}\cdot \frac{(u_i + v_i^2 + 1)^{A_i}(2v_i+1)^{2C_i}}{u_i^{p/2} v_i^q}
\]
\end{proof}

\printbibliography

@article{PaturiPZ99,
  author       = {Ramamohan Paturi and
                  Pavel Pudl{\'{a}}k and
                  Francis Zane},
  title        = {Satisfiability Coding Lemma},
  journal      = {Chic. J. Theor. Comput. Sci.},
  volume       = {1999},
  year         = {1999},
  url          = {http://cjtcs.cs.uchicago.edu/articles/1999/11/contents.html},
  bibsource    = {dblp computer science bibliography, https://dblp.org}
}

@article{PaturiPSZ05,
  author       = {Ramamohan Paturi and
                  Pavel Pudl{\'{a}}k and
                  Michael E. Saks and
                  Francis Zane},
  title        = {An improved exponential-time algorithm for $k$-{SAT}},
  journal      = {J. {ACM}},
  volume       = {52},
  number       = {3},
  pages        = {337--364},
  year         = {2005},
  doi          = {10.1145/1066100.1066101}
}

@inproceedings{Amano23Majority,
  author       = {Kazuyuki Amano},
  editor       = {Satoru Iwata and
                  Naonori Kakimura},
  title        = {Depth-Three Circuits for Inner Product and Majority Functions},
  booktitle    = {34th International Symposium on Algorithms and Computation, {ISAAC}
                  2023, December 3-6, 2023, Kyoto, Japan},
  series       = {LIPIcs},
  volume       = {283},
  pages        = {7:1--7:16},
  publisher    = {Schloss Dagstuhl - Leibniz-Zentrum f{\"{u}}r Informatik},
  year         = {2023},
  url          = {https://doi.org/10.4230/LIPIcs.ISAAC.2023.7},
  doi          = {10.4230/LIPICS.ISAAC.2023.7},
  bibsource    = {dblp computer science bibliography, https://dblp.org}
}

@inproceedings{GKW21circuit,
  author       = {Alexander Golovnev and
                  Alexander S. Kulikov and
                  R. Ryan Williams},
  editor       = {James R. Lee},
  title        = {Circuit Depth Reductions},
  booktitle    = {12th Innovations in Theoretical Computer Science Conference, {ITCS}
                  2021, January 6-8, 2021, Virtual Conference},
  series       = {LIPIcs},
  volume       = {185},
  pages        = {24:1--24:20},
  publisher    = {Schloss Dagstuhl - Leibniz-Zentrum f{\"{u}}r Informatik},
  year         = {2021},
  url          = {https://doi.org/10.4230/LIPIcs.ITCS.2021.24},
  doi          = {10.4230/LIPICS.ITCS.2021.24},
  bibsource    = {dblp computer science bibliography, https://dblp.org}
}

@inproceedings {Valiant77,
    AUTHOR = {Valiant, Leslie G.},
     TITLE = {Graph-theoretic arguments in low-level complexity},
 BOOKTITLE = {Mathematical foundations of computer science ({P}roc. {S}ixth
              {S}ympos., {T}atransk\'{a} {L}omnica, 1977)},
     PAGES = {162--176. Lecture Notes in Comput. Sci., Vol. 53},
      YEAR = {1977},
   MRCLASS = {68A20},
  MRNUMBER = {0660702},
}

@article{HastadJP95,
  author       = {Johan H{\aa}stad and
                  Stasys Jukna and
                  Pavel Pudl{\'{a}}k},
  title        = {Top-Down Lower Bounds for Depth-Three Circuits},
  journal      = {Comput. Complex.},
  volume       = {5},
  number       = {2},
  pages        = {99--112},
  year         = {1995},
  doi          = {10.1007/BF01268140}
}

@inproceedings{FranklGT22,
  author       = {Peter Frankl and
                  Svyatoslav Gryaznov and
                  Navid Talebanfard},
  editor       = {Mark Braverman},
  title        = {A Variant of the {VC}-Dimension with Applications to Depth-3 Circuits},
  booktitle    = {13th Innovations in Theoretical Computer Science Conference, {ITCS}
                  2022, January 31 - February 3, 2022, Berkeley, CA, {USA}},
  series       = {LIPIcs},
  volume       = {215},
  pages        = {72:1--72:19},
  publisher    = {Schloss Dagstuhl - Leibniz-Zentrum f{\"{u}}r Informatik},
  year         = {2022},
  doi          = {10.4230/LIPIcs.ITCS.2022.72}
}

@article{PaturiSZ00,
  author       = {Ramamohan Paturi and
                  Michael E. Saks and
                  Francis Zane},
  title        = {Exponential lower bounds for depth three Boolean circuits},
  journal      = {Comput. Complex.},
  volume       = {9},
  number       = {1},
  pages        = {1--15},
  year         = {2000},
  doi          = {10.1007/PL00001598}
}

@article{GGM24,
  author       = {Mika G{\"{o}}{\"{o}}s and
                  Ziyi Guan and
                  Tiberiu Mosnoi},
  title        = {Depth-3 circuits for inner product},
  journal      = {Inf. Comput.},
  volume       = {300},
  pages        = {105192},
  year         = {2024},
  url          = {https://doi.org/10.1016/j.ic.2024.105192},
  doi          = {10.1016/J.IC.2024.105192},
  bibsource    = {dblp computer science bibliography, https://dblp.org}
}

@article{Dancik96,
  author       = {Vlado Danc{\'{\i}}k},
  title        = {Complexity of Boolean Functions Over Bases with Unbounded Fan-In Gates},
  journal      = {Inf. Process. Lett.},
  volume       = {57},
  number       = {1},
  pages        = {31--34},
  year         = {1996},
  url          = {https://doi.org/10.1016/0020-0190(95)00182-4},
  doi          = {10.1016/0020-0190(95)00182-4},
  bibsource    = {dblp computer science bibliography, https://dblp.org}
}

@book{PWM2024analytic,
  title={Analytic combinatorics in several variables},
  author={Pemantle, Robin and Wilson, Mark C and Melczer, Stephen},
  volume={212},
  year={2024},
  publisher={Cambridge University Press}
}

@book{bv2004convex,
  title={Convex optimization},
  author={Boyd, Stephen and Vandenberghe, Lieven},
  year={2004},
  publisher={Cambridge university press}
}

@book{tc1999infotheory,
  title={Elements of information theory},
  author={Cover, Thomas M},
  year={1999},
  publisher={John Wiley \& Sons}
}

@article{FindGHK23,
  author       = {Magnus Gausdal Find and
                  Alexander Golovnev and
                  Edward A. Hirsch and
                  Alexander S. Kulikov},
  title        = {Improving $3N$ Circuit Complexity Lower Bounds},
  journal      = {Comput. Complex.},
  volume       = {32},
  number       = {2},
  pages        = {13},
  year         = {2023},
  url          = {https://doi.org/10.1007/s00037-023-00246-9},
  doi          = {10.1007/S00037-023-00246-9}
}

@inproceedings{Li022,
  author       = {Jiatu Li and
                  Tianqi Yang},
  title        = {$3.1n - o(n)$ circuit lower bounds for explicit
                  functions},
  booktitle    = {54th Annual {ACM} {SIGACT} Symposium on Theory of Computing, {STOC}},
  pages        = {1180--1193},
  publisher    = {{ACM}},
  year         = {2022},
  url          = {https://doi.org/10.1145/3519935.3519976},
  doi          = {10.1145/3519935.3519976}
}

@inproceedings{Li23,
  author       = {Xin Li},
  title        = {Two Source Extractors for Asymptotically Optimal Entropy, and (Many)
                  More},
  booktitle    = {64th {IEEE} Annual Symposium on Foundations of Computer Science, {FOCS}},
  pages        = {1271--1281},
  publisher    = {{IEEE}},
  year         = {2023},
  url          = {https://doi.org/10.1109/FOCS57990.2023.00075},
  doi          = {10.1109/FOCS57990.2023.00075},
  bibsource    = {dblp computer science bibliography, https://dblp.org}
}

@inproceedings{HuangIV22,
  author       = {Xuangui Huang and
                  Peter Ivanov and
                  Emanuele Viola},
  title        = {Affine Extractors and {AC0}-Parity},
  booktitle    = {Approximation, Randomization, and Combinatorial Optimization. Algorithms
                  and Techniques, {APPROX/RANDOM}},
  series       = {LIPIcs},
  volume       = {245},
  pages        = {9:1--9:14},
  year         = {2022},
  url          = {https://doi.org/10.4230/LIPIcs.APPROX/RANDOM.2022.9},
  doi          = {10.4230/LIPICS.APPROX/RANDOM.2022.9},
  bibsource    = {dblp computer science bibliography, https://dblp.org}
}

@article{IPZ01exponential,
  author       = {Russell Impagliazzo and
                  Ramamohan Paturi and
                  Francis Zane},
  title        = {Which Problems Have Strongly Exponential Complexity?},
  journal      = {J. Comput. Syst. Sci.},
  volume       = {63},
  number       = {4},
  pages        = {512--530},
  year         = {2001},
  url          = {https://doi.org/10.1006/jcss.2001.1774},
  doi          = {10.1006/JCSS.2001.1774},
  bibsource    = {dblp computer science bibliography, https://dblp.org}
}

@inproceedings{Ibex2015,
  title={Global optimization based on contractor programming: An overview of the ibex library},
  author={Ninin, Jordan},
  booktitle={International Conference on Mathematical Aspects of Computer and Information Sciences},
  pages={555--559},
  year={2015},
  organization={Springer}
}

\appendix

\section{Proving coefficient asymptotics}\label{subsec:bivariateOptApproxCauchy}

In this section, we will prove our required lemma regarding bounding coefficients of power series of an analytic function - \cref{lem:bivariateOptApproxCauchy}.

To help prove this, we will require the following well known result that coefficient expansion of a generating function can be expressed as a Cauchy integral (see equation 1.5 from \cite{PWM2024analytic}):
\begin{theorem}[Bivariate Cauchy integral formula]
\label{thm:bivariateCauchyIntegralFormula}
Let $f: \C^2\to \C$ be analytic on a compact neighborhood $\cN$ around $(0, 0)$. Let $f(u, v) = \sum_{r = (r_1, r_2)\in \N^2} C_r u^{r_1}v^{r_2}$ where for all $i\in \N^2$, $C_i \in \R$ and $C_i \ge 0$. Then, for all $r\in \N^2$, it holds that
\[
    C_r = \oint_{x\in \cN} \left(\frac{1}{2\pi i}\right)^2 f(x)\cdot x_1^{-r_1-1}\cdot x_2^{-r_2-1}\, dx. 
\] 
\daniel{maybe simplify $(1/2pi i)^2 to -1/4 (pi^2)$}
\mohit{Hmm I kinda like it this way currently since for dimension $d$, we can just substitute $2$ for $d$ and things are nice.}
\daniel{
is x supposed to be in $\R$ or $\R^2$? f(x) doesn't make sense if in $\R$, and $x^{-r-1}$ is a pain if in $\R^2$. The PWM book has $f$ take inputs in $\C$, but that's not how you have defined your $f$.
Also, are you sure this holds when you just need to be analytic on $\R^2$ rather than on $\C$? It probably does but I don't see it immediately. (The same for future such theorems, only mentioning it once.)
}
\mohit{Thanks for pointing this out, it should have been $\C^2$. I fixed it everywhere. Also changed the statement a bit to make it clear that $x\in \C^2$ and clarified the operation}
\end{theorem}

We will utilize the following result regarding bounding Cauchy integral:

\begin{theorem}[Theorem 5.2 from \cite{PWM2024analytic}, specialized]
\label{thm:bivariateOptApproxCauchy}
Let $A, \phi: \C^2\to \C$ be analytic on a compact neighborhood $\cN$ around $(0, 0)$. Furthermore assume that the real part of $\phi$ is nonnegative on $\cN$ and vanishes only at $(0, 0)$, and that the Hessian matrix $\cH_{\phi}$ of $\phi$ is such that $\cH_{\phi}(0, 0)$ is non-singular. Let, $I(\lambda) = \int_{\cN} A(z) e^{-\lambda \phi(z)}$. Then for $\lambda \le \frac{1}{2}$, we have that:
\[
I(\lambda) 
= A(0, 0)\frac{2\pi}{\sqrt{\det(\cH_{\phi}(0, 0))}}\cdot \frac{1}{\lambda}\left(1 + O\left(\frac{1}{\lambda}\right)\right).
\]
\end{theorem}

Using these, we obtain the following estimate regarding estimating coefficient of an analytic function:

\begin{proof}[Proof of \cref{lem:bivariateOptApproxCauchy}]
We apply \cref{thm:bivariateCauchyIntegralFormula} with the contour $\cC$ being product of circles with $\abs{u} = u_0, \abs{v} = v_0$ so that
\[
    C_{p, q} = \frac{-1}{4\pi^2} \int_{(u, v)\in \cC} f(u, v) u^{-p-1} v^{-q-1} du dv.
\]
For $\theta = (\theta_1, \theta_2)\in [-\pi, \pi]^2$, we perform change of variables to express the points in the contour as $u = u(\theta) = u_0 e^{i \theta_1}, v = v(\theta) = v_0 e^{i \theta_2}$.
We see that $du = u_0 i e^{i\theta_1} d\theta_1$ and $dv = v_0 i e^{i\theta_2} d\theta_2$. 
We also use the fact that $\exp(h(u, v)) = \frac{f(u, v)}{u^p v^q}$ to obtain that
\[
    C_{p, q} = \frac{1}{4\pi^2} \int_{\theta\in [\pi, \pi]^2} \exp(h(u(\theta), v(\theta))) d\theta. %I can't get these numbers, but I'm probably missing something. Might be worth a second step in case other people also miss it though.
\]
Let $\lambda = \sqrt{p^2 + q^2}$. We perform one last change of variables where we let $h'(u, v) = h(u, v) / \lambda$ to express the integral as:
\[
    C_{p, q} = \frac{\exp(h(u_0, v_0))}{4\pi^2} \int_{\theta\in [\pi, \pi]^2} \exp\left(\lambda \cdot \left(h'(u(\theta), v(\theta)) - h'(u_0, v_0)\right)\right) d\theta.
\]

We will apply \cref{thm:bivariateOptApproxCauchy} with $A(\theta) = 1$, $\phi(\theta) = h'(u_0, v_0) - h'(u(\theta), v(\theta))$.
We claim the following:
\begin{claim}
\label{claim:optApproxCaucyHelper}
\[
\det(\cH_{\phi}(0, 0)) = \left(\frac{u_0v_0}{\lambda}\right)^2\det(\cH_{h'}(u_0, v_0)).
\]
\end{claim}
We will prove this at the end and for now will just assume it.  

We first show that we satisfy all of the preconditions laid out by \cref{thm:bivariateOptApproxCauchy}.
First, we have that $\phi(0, 0) = 0$ and by unique maximality of $u_0, v_0$ that for all $\theta\ne (0, 0)$, it follows that $\phi(\theta) > 0$. 
Second, we show that $\cH_\phi(0, 0)$ is non-singular.
By assumption, we know that $\cH_h(u_0, v_0)$ is positive definite and so must be true for $\cH_{h'}(u_0, v_0)$, making it singular.
Therefore by \cref{claim:optApproxCaucyHelper}, it follows that $\cH_\phi(0, 0)$ is non-singular as well.
Lastly, by assumption we have that $\lambda \le \frac{1}{2}$.
Hence, we can indeed apply \cref{thm:bivariateOptApproxCauchy} with the same $\lambda$ to infer that
\[
    C_{p,q} = \frac{\exp(h(u_0, v_0))}{4\pi^2} \cdot \frac{2\pi}{\sqrt{\det({\cH_{\phi}(0, 0)})}}\cdot \frac{1}{\lambda}\cdot \left(1 + O\left(\frac{1}{\lambda}\right)\right).
\]
Using \cref{claim:optApproxCaucyHelper} and the fact that $\exp(h(u_0, v_0)) = \frac{f(u_0, v_0)}{u_0^q v_0^q}$, our desired bound on $C_{p, q}$ follows.

We lastly prove our helper claim that relates $\det(\cH_{\phi}(0, 0))$ and $\det(\cH_{h'}(u_0, v_0))$.
\begin{proof}[Proof of \cref{claim:optApproxCaucyHelper}]
Let $g(\theta) = - h'(u(\theta), v(\theta))$ so that $\phi(\theta) = h'(u_0, v_0) + g(\theta)$.
Since $h'(u_0, v_0)$ is a constant, we have that $\cH_g(0, 0) = \cH_\phi(0, 0)$.
Hence, it suffices to show that $\det(\cH_g(0, 0)) = \left(\frac{u_0v_0}{\lambda}\right)^2\det(\cH_{h'}(u_0, v_0))$.
Using chain rule, we see that for $i\in [2]$:
\[
\frac{\partial g}{\partial \theta_i} = \frac{\partial h'}{\partial u} \frac{\partial u}{\partial \theta_i} + \frac{\partial h'}{\partial v} \frac{\partial v}{\partial \theta_i}
\]
When we evaluate this at $(0, 0)$,  we will evaluate $\frac{\partial h'}{\partial u}$ and $\frac{\partial h'}{\partial v}$ at $(u_0, v_0)$. 
By assumption $\nabla_h(u_0, v_0) = 0$ and so $\nabla_{h'}(u_0, v_0) = 0$.
Using above, this implies that $\nabla_g(0, 0) = 0$.

Using chain rule, the fact that $\nabla_g(0, 0) = 0$, and recalling the definition of $u(\theta)$ and $v(\theta)$, we obtain that 
\[
    \cH_{g}(0, 0) = D \cH_{h'}(u_0, v_0) D
\]
where $D$ is the $2\times 2$ diagonal matrix with entries $iu_0, iv_0$.

With this we finally compute that 
\begin{align*}
\det(\cH_g(0, 0)) 
& = \det(D \cH_{h'}(u_0, v_0) D)\\
& = \det(D)\cdot \det(\cH_{h'}(u_0, v_0))\cdot \det(D)\\
& = (u_0 v_0)^2 \det(\cH_{h'}(u_0, v_0))\\
& = \frac{(u_0 v_0)^2}{\lambda^2} \det(\cH_{h}(u_0, v_0))
\end{align*}
where for the last equality we used the fact that $h' = h / \lambda$, which implies $\cH_{h'}(u_0, v_0) = \frac{1}{\lambda}\cH_{h}(u_0, v_0)$.
\end{proof}
\end{proof}

% \appendix

%\subfile{sections/appendix}

\end{document}